\documentclass[twocolumn,times]{aastex63}

\usepackage{CJKutf8}
\usepackage{amsmath}
\usepackage{amssymb}
\usepackage{amsfonts}
\usepackage{graphicx}
\usepackage{wasysym}
\usepackage{multirow}
\usepackage{mdframed}
\usepackage[T1]{fontenc}

 \usepackage{lineno}

\newcommand{\kms}{\mbox{km\,s$^{-1}$}}

\newcommand{\msol}{\mbox{$M_\odot$}}

\newcommand{\mjypbm}{\mbox{mJy\,beam$^{-1}$}}

\newcommand{\hii}{\mbox{H\,{\sc ii}}}

\newcommand{\methanol}{\mbox{$\rm CH_3OH$}}

\newcommand{\htwoco}{\mbox{$\rm H_2CO$}}

\newcommand{\ammonia}{\mbox{$\rm NH_3$}}
\newcommand{\gcm}{\mbox{g cm$^{-2}$}} 
\newcommand{\lsun}{\mbox{\,$L_\odot$}}
\newcommand{\msun}{\mbox{\,$M_\odot$}}
\newcommand{\msunyr}{\mbox{\,$M_\odot$ yr$^{-1}$}}

\hypersetup{linkcolor=green,citecolor=blue,filecolor=cyan,urlcolor=magenta}
\hypersetup{linkcolor=red,citecolor=blue,filecolor=cyan,urlcolor=magenta}

\received{}
\revised{}
\accepted{}
\submitjournal{ApJ}

\shorttitle{}
\shortauthors{Cheng et al.}

\begin{document}
\begin{CJK}{UTF8}{gbsn}

\title{INvestigations of massive Filaments ANd sTar formation (INFANT). I. Core Identification and Core Mass Function}

\correspondingauthor{Yu Cheng, Xing Lu}
\email{ycheng.astro@gmail.com, xinglu@shao.ac.cn}

\author[0000-0002-0786-7307]{Yu Cheng (程宇)}
\affil{National Astronomical Observatory of Japan, 2-21-1 Osawa, Mitaka, Tokyo 181-8588, Japan}

\author[0000-0003-2619-9305]{Xing Lu (吕行)}
\affil{Shanghai Astronomical Observatory, Chinese Academy of Sciences, 80 Nandan Road, Shanghai 200030, People's Republic of China}

\author{Patricio Sanhueza}
\affil{National Astronomical Observatory of Japan, National Institutes of Natural Sciences, 2-21-1 Osawa, Mitaka, Tokyo 181-8588, Japan}

\author[0000-0003-2300-2626]{Hauyu Baobab Liu}
\affil{Physics Department, National Sun Yat-Sen University, No. 70, Lien-Hai Road, Kaohsiung City 80424, Taiwan, R.O.C.}

 \author{Qizhou Zhang}
 \affil{Center for Astrophysics | Harvard \& Smithsonian, 60 Garden Street, Cambridge, MA 02138, USA}

\author[0000-0003-1480-4643]{Roberto Galv\'an-Madrid}
\affil{Instituto de Radioastronom\'ia y Astrof\'isica, Universidad Nacional Aut\'onoma de M\'exico, Morelia, Michoac\'an 58089, M\'exico.}

\author{Ke Wang}
\affil{Kavli Institute for Astronomy and Astrophysics, Peking University, 5 Yiheyuan Road, Haidian District, Beijing 100871, Peopleʼs Republic of China}

\author{Fumitaka Nakamura} 
\affil{National Astronomical Observatory of Japan, National Institutes of Natural Sciences, 2-21-1 Osawa, Mitaka, Tokyo 181-8588, Japan}

\author{Tie Liu}
\affil{Shanghai Astronomical Observatory, Chinese Academy of Sciences, 80 Nandan Road, Shanghai 200030, People's Republic of China}

\author{Siyi Feng}
\affil{Department of Astronomy, Xiamen University, Zengcuo'an West Road, Xiamen, 361005, People's Republic of China}

\author{Shanghuo Li}
\affil{Max Planck Institute for Astronomy, Konigstuhl 17, D-69117 Heidelberg, Germany}

\author{Sihan Jiao}
\affil{National Astronomical Observatories, Chinese Academy of Sciences, Beijing 100101, Peopleʼs Republic of China}

\author{Kei E. I. Tanaka}
\affil{Department of Earth and Planetary Sciences, Tokyo Institute of Technology, Meguro, Tokyo, 152-8551, Japan}

\author{Xunchuan Liu}
\affil{Shanghai Astronomical Observatory, Chinese Academy of Sciences, 80 Nandan Road, Shanghai 200030, People's Republic of China}

\author{Pak Shing Li}
\affil{Shanghai Astronomical Observatory, Chinese Academy of Sciences, 80 Nandan Road, Shanghai 200030, People's Republic of China}

\author{Qiuyi Luo}
\affil{Shanghai Astronomical Observatory, Chinese Academy of Sciences, 80 Nandan Road, Shanghai 200030, People's Republic of China}

\author{Qilao Gu}
\affil{Shanghai Astronomical Observatory, Chinese Academy of Sciences, 80 Nandan Road, Shanghai 200030, People's Republic of China}

\author{Yuxin Lin}
\affil{Max-Planck-Institut f\"ur Extraterrestrische Physik, Giessenbachstr. 1, D-85748 Garching bei M\"unchen, Germany}

\author{Andr\'es E.\ Guzm\'an}
\affil{Joint Alma Observatory (JAO), Alonso de C\'ordova 3107, Vitacura, Santiago, Chile}




\begin{abstract}

Filamentary structures are ubiquitously found in high-mass star-forming clouds. To investigate the relationship between filaments and star formation, we carry out the INFANT (INvestigations of massive Filaments ANd sTar formation) survey, a multi-scale, multi-wavelength survey of massive filamentary clouds with ALMA band 3/band 6 and VLA K band. In this first paper, we present the ALMA band 6 continuum observations toward a sample of 8 high-mass star forming filaments. We covered each target with approximately rectangular mosaic field of view with two 12-m array configurations, achieving an angular resolution of $\sim$0\farcs{6} (2700 AU at 4.5 kpc) and a continuum rms of $\sim$0.1 \mjypbm ($\sim$0.06 \msol{} in gas mass assuming 15 K). We identify cores using the {\it getsf} and {\it astrodendro} and find the former is more robust in terms of both identification and measuring flux densities. We identify in total 183 dense cores (15--36 cores in each cloud) and classify their star formation states via outflow and warm gas tracers. The protostellar cores are statistically more massive than the prestellar cores, possibly indicating further accretion onto cores after formation of protostars. For the high-mass end ($M_\text{core}$ $>$ 1.5 \msun) of the core mass function (CMF) we derive a power-law index of $-$1.15 $\pm$ 0.12 for the whole sample, and $-$1.70 $\pm$ 0.25 for the prestellar population. We also find a steepening trend in CMF with cloud evolution ($-$0.89 $\pm$ 0.15 for the young group v.s.\ $-$1.44 $\pm$ 0.25 for the evolved group) and discuss its implication for cluster formation.

\end{abstract}

\section{Introduction}\label{sec:intro}

Filamentary structures pervade the interstellar medium over a wide range of scales \citep{Hacar23}. Within star-forming molecular gas, the presence of parsec-scale filamentary structures and their potential importance for star formation have been recognized for decades. Observations towards nearby clouds such as Taurus and Orion revealed prominent filamentary structures in spectral line and dust continuum emission \citep[e.g.,][]{Bally87,Abergel94}. The Galactic wide surveys carried out with the {\it Herschel} Space Observatory have demonstrated the ubiquity of filaments throughout the Galactic plane \citep{Molinari10}, and facilitated a detailed characterization of filaments in nearby clouds with characteristic length of a few parsecs and a width of $\sim$0.1~pc when observed at resolutions of the {\it Herschel}  Space Telescope \citep{Andre10,Arzoumanian11,Arzoumanian19,Konyves20}. When observed at high enough resolutions, these filaments are often shown to contain fibrous substructures, which are velocity coherent ``fibres'' with sub-parsec lengths, small characteristic widths of 0.02--0.1~pc, and (tran-)sonic velocity dispersions \citep{Hacar13,Hacar18,Fernandez-Lopez14,Schmiedeke21,Li22}.

The relationship between filaments and star formation is still being intensively studied \citep[][and references therein]{Andre14,Hacar23}. In nearby clouds filaments make up more than 80\% of the dense gas mass at $N_{H_2}$ $\rm >7\times 10^{21} cm^{-2}$  \citep{Arzoumanian19}, and harbor most prestellar cores \citep{Konyves15,Konyves20}. Fragmentation in filaments set the initial condition for the formation and evolution of cores, and may potentially regulate their mass distributions \citep{Andre19}. 
Based on { the Hi-GAL project \citep{Molinari10}}, \citet{Schisano14} found that cores with surface densities in excess of the expected critical values for high-mass star formation are only found on the filaments. The conjunction of several filaments, or a ``hub'', where the density can be greatly enhanced, have been frequently observed to be the formation sites of high-mass stars or young clusters \citep{Myers09,Galvn10,Liu12a,Liu12,Kumar20}. Many hub-filament systems appear to be highly dynamic, with converging longitude flows feeding their central hubs at a high accretion rate of order $10^{-4}$--$10^{-3}$ \msunyr \citep{Kirk13,Peretto13,Peretto14,Hacar17,Chen19,Sanhueza21,Zhou22}.


While a significant step forward, our detailed knowledge of internal filament properties is still largely limited to the solar neighborhood. Systematic surveys to characterize massive filaments in distant high-mass star forming clouds are still rare. These filaments are likely different from their low-mass counterparts \citep[e.g., $\gtrsim$5 times larger mass per unit length, stronger turbulence;][]{Lu18}. Although a large number of high-mass star forming filaments have been catalogued \citep[e.g.,][]{Carey00,Wang16}, high angular resolution studies that are able to resolve them down to dense cores scales $\lesssim$0.02 pc are mostly case studies \citep[e.g.,][]{Wang14,Zhang15,Li20}. In light of this we carried out the INFANT (INvestigations of massive Filaments ANd sTar formation) survey, a multi-wavelength, multi-scale survey of 8 massive filamentary clouds with the Atacama Large Millimeter/submillimeter Array (ALMA) band~3/band~6 and Very Large Array (VLA) K~band, aimed at investigating the relationship between filaments and star formation across a range of evolutionary stages. This first paper will focus on the ALMA band~6 continuum and properties of dense cores. The paper is organized as follows: a description of the sample and observations is given in \autoref{sec:sample} and \autoref{sec:observation}, respectively; we then present the continuum map and core identification and characterization results in \autoref{sec:results}, followed by a comparison of the core extraction methods in \autoref{sec:core_iden_comp}.
Further discussions are made in \autoref{sec:discussion}; and a summary is given in \autoref{sec:summary}.


%



\startlongtable
\begin{deluxetable*}{cccccccc}
\tabletypesize{\scriptsize}
\renewcommand{\arraystretch}{1.0}
\tablecaption{Summary of the INFANT sample\label{table:info_sample}}
\tablehead{
\colhead{Target} & \colhead{R.A.} & \colhead{Decl.}  & \colhead{Distance\tablenotemark{a}} & \colhead{$V_{\rm sys}$\tablenotemark{a}} & \colhead{Luminosity\tablenotemark{b}} & \colhead{Mass\tablenotemark{b}} & \colhead{L/M\tablenotemark{b}} \\
\colhead{} & \colhead{(J2000)} & \colhead{(J2000)}  & \colhead{(kpc)} & \colhead{(\kms)} & \colhead{($10^4L_\odot$)} & \colhead{($10^3$\msun)} & \colhead{($L_\odot$/\msun)}
}
\startdata
IRAS 18308$-$0841 & 18:33:33.30  &  $-$08:38:57.0 & 4.6 	& 77 & $2.5_{-0.9}^{+0.7}$ &  $2.7_{-0.7}^{+0.5}$ &  $9.3_{-0.7}^{+0.8}$\\
IRAS 18310$-$0825 & 18:33:47.60  &  $-$08:23:50.0 & 4.8 	& 84 & $2.9_{-1.1}^{+0.6}$ &  $2.5_{-0.8}^{+0.4}$ &  $11.6_{-0.3}^{+0.4}$ \\
IRAS 18337$-$0743 & 18:36:41.00  &  $-$07:39:27.0 & 3.8 	& 60 & $1.2_{-0.4}^{+0.3}$ &  $2.5_{-0.5}^{+0.4}$ &  $4.9_{-0.6}^{+0.5}$ \\
IRAS 18460$-$0307 & 18:48:38.30  &  $-$03:03:53.0 & 4.8 	& 85 & $1.1_{-0.3}^{+0.2}$ &  $1.5_{-0.4}^{+0.4}$ &  $7.6_{-0.2}^{+0.2}$ \\
IRAS 18530$+$0307 & 18:55:32.70  &  $+$02:19:03.0 & 4.6 	& 76 & $2.9_{-0.6}^{+0.6}$ &  $2.6_{-0.7}^{+0.5}$ &  $11.2_{-0.1}^{+0.5}$ \\
IRAS 19074$+$0752 & 19:09:54.00  &  $+$07:57:15.0 & 3.8 	& 56 & $0.7_{-0.1}^{+0.1}$ &  $0.6_{-0.2}^{+0.1}$ &  $11.4_{-0.5}^{+0.8}$ \\
IRAS 19220$+$1432 & 19:24:21.00  &  $+$14:38:08.0 & 5.4 	& 70 & $1.7_{-0.3}^{+0.3}$ &  $1.3_{-0.2}^{+0.2}$ &  $12.7_{-0.1}^{+0.3}$ \\
IRAS 19368$+$2239 & 19:38:58.20  &  $+$22:46:44.0 & 4.4 & 37 & $0.3_{-0.1}^{+0.1}$ &  $1.6_{-0.4}^{+0.3}$ &  $2.1_{-0.1}^{+0.1}$ \\
\enddata
\tablenotetext{a}{Taken from \citet{Lu18}. }
\tablenotetext{b}{Estimated from {\it Herschel} SED fitting. The uncertainties refer to the variation when the region selected for calculation varies by 30\% in area.}
\end{deluxetable*}

\section{Sample}\label{sec:sample}

The eight clouds in this survey have been previously studied with the Submillimeter Array (SMA)  \citep{Lu18}. They are carefully selected from a VLA \ammonia{} survey toward 62 high-mass star-forming regions in \citet{Lu14} with the following criteria \citep[see also][]{Lu18}:

\begin{enumerate}
\item The VLA \ammonia{} emission exhibits filamentary structures with aspect ratios of $>$5 in the moment~0 map. 
\item They are luminous ($\gtrsim$ 5$\times$10$^3$\lsun) and massive ($\gtrsim$ 10$^3$\msun), with sufficient mass reservoir for high-mass star formation. The only exception is IRAS~19074$+$0752, which is associated with an \hii{} region and its molecular gas has been largely consumed or dispersed, leading to a smaller mass. It is included here as a representative for a protocluster in a much more evolved environment.
\item The distances of the eight targets range from 3.8 to 5.4~kpc, thus enabling a relatively uniform spatial resolutions in a single observational setup.
\end{enumerate}

The properties of the targets are listed in \autoref{table:info_sample}. We take the distance and systematic velocity from \citet[][see also a discussion for the uncertainties therein]{Lu18}. We revised the estimation of masses and luminosities based on spectral energy distribution (SED) fitting of {\it Herschel} data from 70~$\mu$m to 500~$\mu$m, as described in \autoref{sec:app_sed}. We find that the resulting masses or luminosities largely depend on how the region is defined. Thus we give a typical range by calculating the limits when the region area varies by 30\%. For the fiducial case we adopt the boundary defined by the primary beam correction equal to 0.5 in our ALMA band~6 observations, which are designed to cover the \ammonia{} emission and SMA 1.3~mm continuum in \citet{Lu14} and \citet{Lu18}.

\citet{Lu18} identified and characterized 50 massive dense cores in these clouds with the SMA and revealed a strong relationship between filaments and high-mass star formation, through (i) filamentary fragmentation in very early evolutionary phases to form dense cores, (ii) accretion flows along filaments that are important for the growth of dense cores and protostars, and (iii) enhancement of nonthermal motion in the filaments by the feedback or accretion during star formation. Nevertheless, the SMA observations had relatively limited spatial resolution ($\sim$3\arcsec) and sensitivity ($\sim$1~\mjypbm) and hence were unable to probe low to intermediate mass dense cores ($\lesssim$5~\msun) and further fragmentation on $\lesssim$0.05 pc scales. 

\section{Observations}\label{sec:observation}

\subsection{ALMA band~6 observations}

The observations were conducted with ALMA in Cycle 5 (Project ID 2017.1.00526.S, PI: Lu, X.), over a period from 2018 March to 2018 August. All the targets were observed at band~6 using two 12-m array configurations, i.e., C43-1 and C43-4. This combination allows us to achieve a spatial resolution of $\sim$0\farcs{6} (2700~au at a typical distance of 4.5~kpc) and the maximum recoverable angular scales of 12\farcs{4} ($\sim$0.27~pc at 4.5~kpc). We covered each cloud using 30 to 40 mosaic fields in the C43-1 and C43-4 array configurations, respectively. For each source the on source time is at least 8 minutes in C43-1 and 20 minutes in C43-4. The detailed information, including the baseline ranges, numbers of effective antennas, and weather conditions, is listed in \autoref{table:info_obs}.

We set the central frequency of the four correlator sidebands to be 218.500, 217.100, 230.438, 231.420~GHz, respectively. Each baseband has a bandwidth of 937.5~MHz and a uniform spectral resolution of 0.488~MHz. This allows for an aggregate bandwidth of 3.75~GHz and also covers lines including $\rm H_2CO~(3_{0,3}-2_{0,2})$, $\rm H_2CO~(3_{2,2}-2_{2,1})$, $\rm H_2CO~(3_{2,1}-2_{2,0})$, $\rm SiO~(5-4)$, $\rm CO~(2-1)$, and $\rm N_2D^+~(3-2)$, with a velocity resolution of 0.63 -- 0.67~\kms. 

The raw data were calibrated with the data reduction pipeline using {CASA 5.4.0} \citep{Mcmullin07}. We constructed the continuum visibility with all line-free channels. The effecitve aggregated continuum bandwidth is typically 70\%--85\% of our overall bandwidth for different sources. The imaging was carried out using {CASA 6.5.0}. We used the tclean task in CASA, with Briggs weighting and a robust parameter of 0.5, and a multiscale algorithm with scales of [0, 5, 15, 50] and a pixel size of 0\farcs{15}. The resultant synthesized beam size and root mean square (rms) noise of the 1.3~mm continuum for each target are listed in \autoref{table:info_cont}. The angular resolutions are 0\farcs{5} -- 0\farcs{7}, while the rms noise ranges from 0.06 to 0.12~\mjypbm.

\startlongtable
\begin{deluxetable*}{cccccccc}
\tabletypesize{\scriptsize}
\renewcommand{\arraystretch}{1.0}
\tablecaption{Summary of the ALMA band~6 observations\label{table:info_obs}}
\tablehead{
\colhead{Schedule block} & \colhead{Sources} & \colhead{Configuration}  & \colhead{Execution date} & \colhead{Number of effective antennas} & \colhead{Baseline} & \colhead{PWV(mm)} & \colhead{Time on source (min)\tablenotemark{a}}
}
\startdata
I18308\_TM1  & I18308, I18310, I18337 & C43-4     &  2018-03-13    & 40 & 14m--783m    & 1.79  & 37\\
            &                       &           &  2018-03-16    & 41 & 14m--783m    & 1.92  & 37 \\ 
I18308\_TM2  & I18308, I18310, I18337 & C43-1     &  2018-06-09    &  41  & 14m--313m    & 1.02  &  37\\
I19220\_TM1  & I19220, I19368                & C43-4     &  2018-03-11    & 41 & 14m--919m    & 0.57  & 37    \\
            &                       &           &  2018-03-11    & 42 & 14m--783m    & 1.93  & 37  \\ 
            &                       &           &  2018-08-28    & 44 & 14m--781m    & 1.16  & 37    \\ 
            &                       &           &  2018-08-29    & 44 & 14m--781m    & 1.39  & 37  \\ 
I19220\_TM2  & I19220, I19368                & C43-1     &  2018-06-09    & 48 & 14m--313m    & 1.11  & 20\\
I19074\_TM1  & I19074                & C43-4     &  2018-03-22    & 46 & 14m--783m    & 1.03  & 22  \\
I19074\_TM2  & I19074                & C43-1     &  2018-06-06    & 45 & 14m--313m    & 0.68  & 8   \\
I18460\_TM1  & I18460, I18530                & C43-4     &  2018-03-16    & 41 & 14m--783m    & 1.80  & 40  \\
I18460\_TM2  & I18460, I18530                & C43-1     &  2018-06-09    & 49 & 14m--313m    & 1.14  & 20  \\
\enddata
\tablenotetext{a}{Time on source is the time per exception block (EB) that is distributed among the different science targets in the EB.}
\end{deluxetable*}

\subsection{VLA K band observations}
The VLA observations were performed in the C and D configurations, respectively, under the project codes of 20A-405 and 21A-210 (PI: Y.\ Cheng). The correlator setup included 16 128-MHz-wide spectral windows to cover continuum emission with a total bandwidth of 2~GHz, as well as narrow spectral windows to cover the metastable \ammonia{} lines from ($J$,$K$) = (1,1) to (5,5), the H$_2$O maser line, four non-metasable \ammonia{} lines, as well as several other lines that will be discussed in a future publication. Here we only focus on the \ammonia{} ($J$,$K$) = (1,1) and (2,2) lines.

The data were manually calibrated using CASA 5.6.1 and 6.2.0. Continuum baseline was estimated by fitting a linear function to the visibility data including only line-free channels along the frequency axis. The data were also binned by two channels, to achieve a velocity resolution of 0.4~\kms{}. Then, the two datasets from the two array configurations were concatenated and imaged, using CASA 5.7.2. We used the \textit{tclean} task in CASA, with Briggs weighting and a robust parameter of 0.5, and a multiscale algorithm with scales of [0, 5, 15, 50, 150] and a pixel size of 0\farcs{2}. The synthesized beam sizes are different among targets and lines, but are 1\farcs{2} $\times$ 1\farcs{0} in general. The rms noise in the 0.4~\kms{} channel ranges from 1.3 to 1.5~mJy\,beam$^{-1}$ (2.4--2.7~K).

\startlongtable
\begin{deluxetable*}{ccccccccccc}
\tabletypesize{\scriptsize}
\renewcommand{\arraystretch}{1.0}
\tablecaption{Summary of the 1.3~mm continuum and core detection\label{table:info_cont}}
\tablehead{
\colhead{Target} & \colhead{Synthesized beam} & \colhead{Physical resolution} & \colhead{RMS} & \colhead{Mass sensitivity} & \colhead{$M_{\rm complete}$} & \colhead{$M_{\rm max}$} &\colhead{$N_{\rm core}$}& \colhead{$N_{\rm pre}$}& \colhead{$N_{\rm proto}$}& \colhead{$N_{\rm outflow}$} \\
\colhead{} & \colhead{(\arcsec$\times$\arcsec)} & \colhead{(AU)} & \colhead{(\mjypbm)} & \colhead{(\msun)} & \colhead{\msun}&\colhead{\msun} & & 
}
\startdata
I18308 & 0.71$\times$0.59 & 2840 & 0.12 & 0.08 & 0.89& 37.6&22  &  7  &  15  &  9\\
I18310 & 0.71$\times$0.60 & 2870 & 0.10 & 0.06 & 0.73& 9.7 & 27  &  19  &  8  &  2 \\
I18337 & 0.72$\times$0.59 & 2870 & 0.12 & 0.07 & 0.52& 17.0 & 36  &  21  &  15  &  10\\
I18460 & 0.74$\times$0.67 & 3110 & 0.11 & 0.07 & 0.71& 9.3 & 15  &  9  &  6  &  4\\
I18530 & 0.75$\times$0.71 & 3210 & 0.12 & 0.08 & 0.79& 12.9 & 35  &  21  &  14  &  9\\
I19074 & 0.60$\times$0.58 & 2600 & 0.08 & 0.05 & 0.53& 4.8 & 15  &  9  &  6  &  3\\
I19220 & 0.53$\times$0.51 & 2300 & 0.06 & 0.04 & 0.79& 4.2 & 17  &  13  &  4  &  3\\
I19368 & 0.68$\times$0.58 & 2770 & 0.10 & 0.06 & 0.71& 6.8 & 16  &  9  &  7  &  3\\
\enddata
\tablecomments{ The mass sensitiviey is calculated assuming a temperature of 15~K. $M_{\rm complete}$ refers to the mass for which the detection for a point source is 90\% complete as estimated by the artificial core insertion experiment. $M_{\rm max}$ is the maxium core mass. $N_{\rm outflow}$ refers to the number of cores that are associated with molecular outflows. $N_{\rm proto}$ refers to the number of cores that are associated with molecular outflows or compact feature in warm core tracers (including \htwoco{} ($3_{2,2}-2_{2,1}$), ($3_{2,1}-2_{2,0}$) or \methanol{} ($4_{2,2}-3_{1,2}$). $N_{\rm pre}$ indicates number of cores that are absent in both outflows and warm core tracers. }
\end{deluxetable*}

\begin{figure*}[ht!]
\centering
\includegraphics[width=0.9\textwidth]{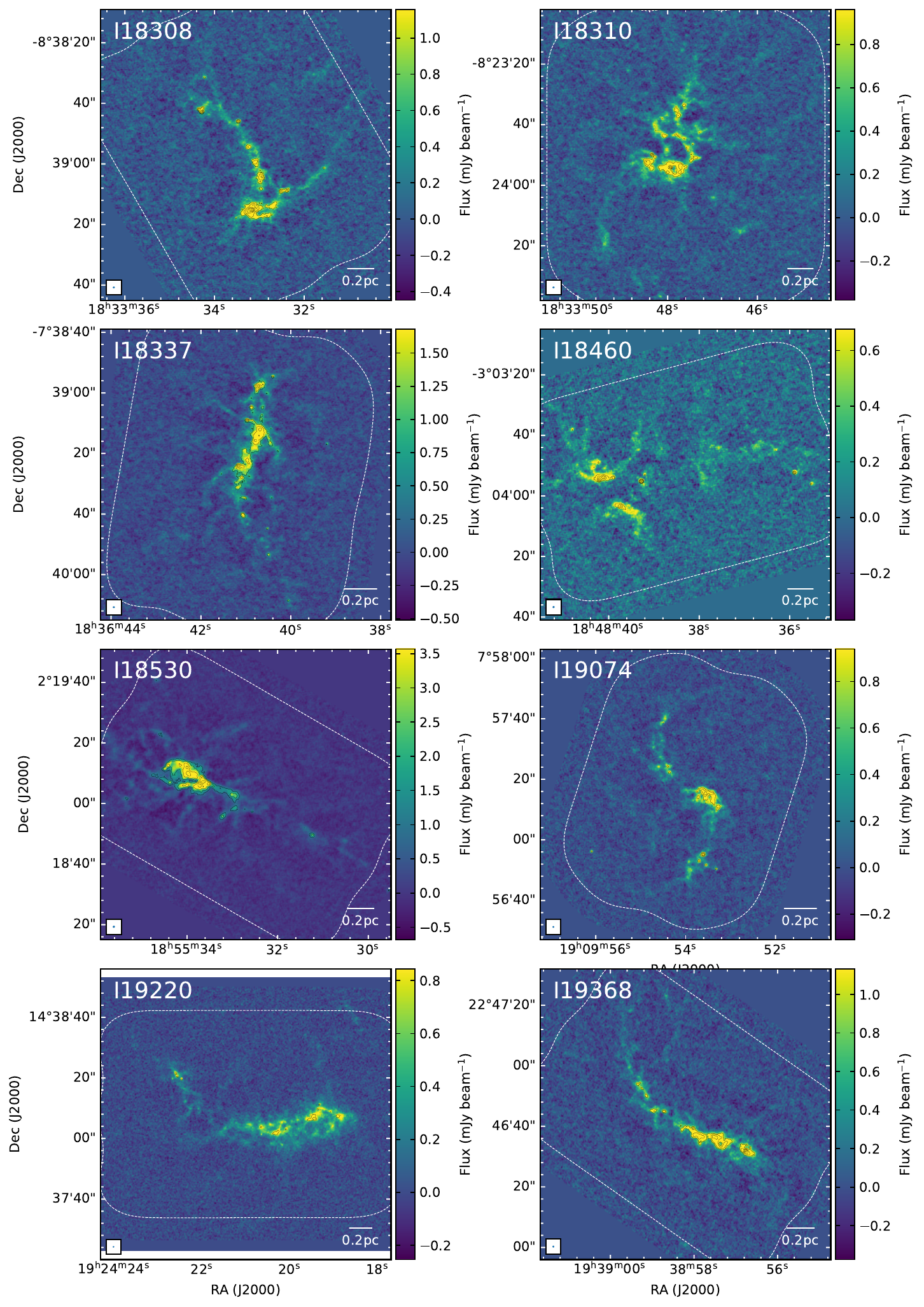}
\caption{Overview of the eight clouds in our sample. ALMA 1.3~mm continuum emission (without primary beam correction) is shown in colorscale and black contours. The contour levels are 0.2~\mjypbm $\times$(5, 10, 20, 40, 80, 160). The dashed loop in each panel shows the ALMA mosaic field.}\label{fig:cont}
\end{figure*}

\section{Results}\label{sec:results}
\subsection{Dust Continuum Emission}

In \autoref{fig:cont} we present the ALMA 1.3~mm continuum maps of the INFANT sample. Overall ALMA recovers 18\%--36\% of the total flux densities measured with IRAM~30m telescope/MAMBO bolometer array at 1.2~mm \citep{Beuther02}. We have scaled the MAMBO fluxes by $(\nu_{\rm eff,ALMA}/\nu_{\rm eff,MAMBO})^4$ to account for the slightly different effective frequencies, i.e., assuming optically thin thermal emission and an opacity law for the dust emission of $\kappa \sim v^2$ \citep{Ossenkopf94}. This fraction reflects different levels of dense gas concentration at smaller scales ($<$12\arcsec, or $\sim$0.25~pc) among the sample.

As discussed in \autoref{sec:sample}, the sample is selected based on an extended morphology in \ammonia{} moment~0 maps. These sources appear to retain similar filamentary emission in the high resolution ALMA 1.3~mm maps, e.g., I18308, I18337, and I19368 being the best examples. In some cases the dust continuum emission does not present prominent filamentary morphology as seen in \ammonia{} in \citet{Lu18}, likely due to spatial filtering effect. For example, I18460 appears as a couple of sparsely distributed dense cores, although these cores are still embedded and connected with relatively diffuse emissions. 

\subsection{Core Identification}\label{sec:core_iden}

\begin{figure*}[ht!]
\epsscale{1.1}\plotone{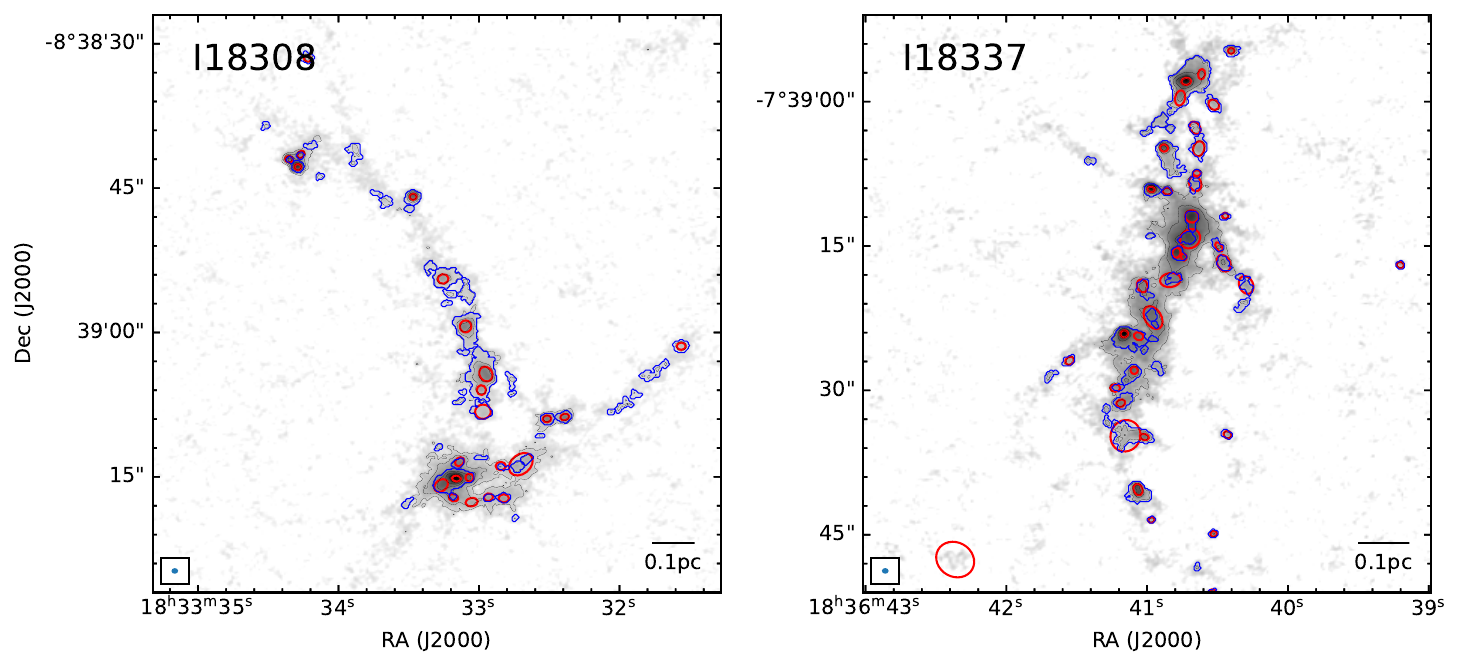}
\caption{Examples of core identification results with {\it getsf} and {\it astrodendro}.
The 1.3~mm image is shown in grey colorscale and black contours with levels of 0.2~\mjypbm $\times$(5, 10, 20, 40, 80, 160). Red ellipses show the FWHM size of cores extracted by {\it getsf}, and blue contours show the core boundaries identified by {\it astrodendro}. }\label{fig:map_iden}
\end{figure*}

\begin{figure*}[ht!]
\epsscale{0.8}\plotone{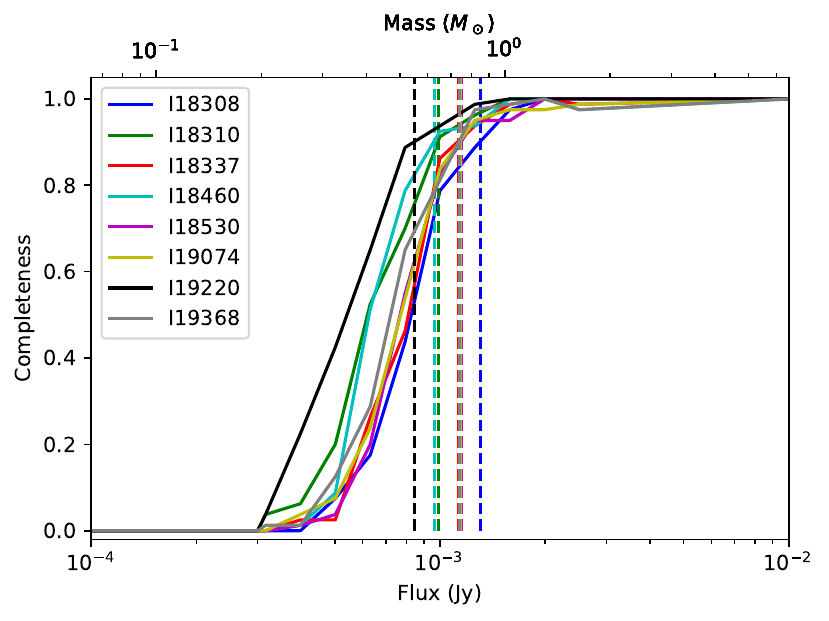}
\caption{Recovery fraction (or completeness level) as a function of flux density for different targets in the sample. The vertical lines denote the fluxes for which the detections are 90\% complete. The corresponding mass is calculated assuming a distance of 4.5~kpc, a dust temperature of 15~K and a gas-to-dust-mass ratio of 100.}\label{fig:complete}
\end{figure*}

We employ the {\it getsf} method to identify and extract properties of dense cores from the 1.3~mm continuum images. A comparison of the performance of {\it getsf} and another commonly used algorithm {\it astrodendro} is further discussed in \autoref{sec:core_iden_comp}.
{\it Getsf} is a method for extracting sources and filaments in astronomical images based on separation of multiscale structural components \citep{Men21}. In this algorithm the relatively round sources and elongated filaments are separated from each other and from the background. In this work we focus on dense cores, i.e., the sources returned by {\it getsf}. 
This method has a single free parameter, i.e., the maximum size of the sources to be extracted, for which we adopt 10\arcsec, i.e., $\sim$0.2~pc at a typical distance of 4.5~kpc. The core identification is based on maps that are not corrected for primary beam response; the fluxes returned by {\it getsf} are corrected by the beam response factor for further analysis.

\autoref{fig:map_iden} shows examples of the identification results towards I18308 and I18337. {\it Getsf} returns an estimation of source sizes and flux densities, as listed in \autoref{table:info_core}. The final catalogue includes 183 cores that pass the recommended optimal selection criteria based on benchmark tests (equation 1 in \citealt{Men21b}), including an aspect ratio lower than 2, peak and integrated flux signal-to-noise ratios above 2, etc. The core numbers for different targets in our sample range from 15 to 36, with a median of 21. I18460 and I19074 has the smallest number of cores (15), while I18337 has the largest core number of 36. 

To estimate the completeness level, we run experiments of artificial core insertion by generating artificial cores of certain flux densities, randomly putting them in the original image, and running {\it getsf} again to check whether they can be picked out. For simplicity, the artificial cores are assumed to have the same shape as the synthesized beam, i.e., the limiting case appropriate for small, unresolved cores. In each experiment, we insert 10 cores of a given total flux, and repeat this for 10 times. The results are shown in \autoref{fig:complete}. The 90\% completeness levels range from 0.8 to 1.3~mJy, i.e.,  around 10 $\times$ overall image rms as listed in \autoref{table:info_cont}.

\subsection{Properties of dense cores}

\startlongtable
\begin{deluxetable*}{cccccccccccc}
\tabletypesize{\scriptsize}
\renewcommand{\arraystretch}{1.0}
\tablecaption{Core properties extracted by {\it getsf}\label{table:info_core}}
\tablehead{
\colhead{Region} & \colhead{Index} & \colhead{$\alpha$(J2000)} & \colhead{$\delta$(J2000)}  & \colhead{$\rm FWHM$} & \colhead{PA} & \colhead{Flux Density} & \colhead{Mass} & \colhead{$T_{\rm NH_3}$}& \colhead{$\sigma_{\rm NH_3}$} & \colhead{outflow} & \colhead{pre/proto}\\
\colhead{} & \colhead{} & \colhead{(hh:mm:ss)} & \colhead{(dd:mm:ss)}  & \colhead{(\arcsec$\times$\arcsec)} & \colhead{($^{\circ}$)} & \colhead{(mJy)} & \colhead{(\msun)} & \colhead{(K)} & \colhead{(\kms)} & &}
\startdata
I18308 & 1 & 18:33:33.16 & $-$08:39:15.0 & 0.96$\times$0.64 & 84 & 90.7 & 37.61 & 21.7$\pm$0.9 & 1.07$\pm$0.04 & outflow & protostellar \\
I18308 & 2 & 18:33:34.29 & $-$08:38:43.0 & 0.74$\times$0.63 & 90 & 25.3 & 24.81 & 11.7$\pm$0.8 & 0.57$\pm$0.04 & outflow & protostellar \\
I18308 & 3 & 18:33:33.27 & $-$08:39:16.0 & 1.58$\times$1.17 & 127 & 20.1 & 9.40 & 19.8$\pm$0.6 & 0.82$\pm$0.03 & no outflow & protostellar \\
I18308 & 4 & 18:33:32.95 & $-$08:39:04.0 & 1.59$\times$1.33 & 31 & 16.1 & 13.19 & 13.2$\pm$0.6 & 0.54$\pm$0.03 & no outflow & protostellar \\
I18308 & 5 & 18:33:33.10 & $-$08:38:59.0 & 1.24$\times$1.21 & 177 & 11.3 & 10.69 & 12.0$\pm$0.4 & 0.45$\pm$0.03 & outflow & protostellar \\
I18308 & 6 & 18:33:33.47 & $-$08:38:46.0 & 0.74$\times$0.64 & 90 & 9.9 & 5.76 & - & - & outflow & protostellar \\
I18308 & 7 & 18:33:32.70 & $-$08:39:14.0 & 2.79$\times$2.02 & 128 & 9.8 & 5.85 & 16.5$\pm$0.7 & 0.62$\pm$0.03 & no outflow & prestellar \\
I18308 & 8 & 18:33:34.35 & $-$08:38:42.0 & 0.86$\times$0.68 & 82 & 7.9 & 4.58 & - & - & no outflow & protostellar \\
I18308 & 9 & 18:33:33.07 & $-$08:39:15.0 & 0.88$\times$0.82 & 58 & 6.9 & 2.98 & 21.1$\pm$1.0 & 1.02$\pm$0.04 & outflow & protostellar \\
I18308 & 10 & 18:33:32.52 & $-$08:39:09.0 & 0.91$\times$0.72 & 97 & 6.1 & 2.35 & 23.0$\pm$1.6 & 0.67$\pm$0.04 & outflow & protostellar \\
I18308 & 11 & 18:33:32.39 & $-$08:39:09.0 & 0.94$\times$0.69 & 103 & 6.0 & 3.01 & 18.7$\pm$1.1 & 0.36$\pm$0.02 & outflow & protostellar \\
I18308 & 12 & 18:33:33.14 & $-$08:39:13.0 & 1.07$\times$0.84 & 130 & 5.5 & 2.66 & 19.2$\pm$0.9 & 0.84$\pm$0.04 & no outflow & prestellar \\
I18308 & 13 & 18:33:33.26 & $-$08:38:54.0 & 1.17$\times$1.00 & 100 & 4.8 & 5.31 & 10.9$\pm$0.4 & 0.28$\pm$0.02 & no outflow & prestellar \\
I18308 & 14 & 18:33:32.97 & $-$08:39:08.0 & 1.67$\times$1.49 & 95 & 4.4 & 2.88 & 15.3$\pm$0.9 & 0.43$\pm$0.03 & no outflow & prestellar \\
I18308 & 15 & 18:33:34.27 & $-$08:38:41.0 & 0.88$\times$0.67 & 134 & 3.9 & 3.68 & 11.9$\pm$0.8 & 0.57$\pm$0.04 & outflow & protostellar \\
I18308 & 16 & 18:33:33.18 & $-$08:39:17.0 & 1.04$\times$0.76 & 74 & 3.7 & 2.13 & 16.8$\pm$0.7 & 0.80$\pm$0.03 & no outflow & prestellar \\
I18308 & 17 & 18:33:32.82 & $-$08:39:17.0 & 1.10$\times$0.86 & 89 & 3.0 & 1.45 & 19.0$\pm$1.3 & 0.47$\pm$0.03 & no outflow & protostellar \\
I18308 & 18 & 18:33:32.98 & $-$08:39:06.0 & 1.03$\times$0.98 & 7 & 2.8 & 2.41 & 12.9$\pm$0.6 & 0.58$\pm$0.03 & no outflow & prestellar \\
I18308 & 19 & 18:33:32.93 & $-$08:39:17.0 & 0.99$\times$0.81 & 103 & 2.8 & 1.01 & 24.4$\pm$1.8 & 0.60$\pm$0.04 & no outflow & protostellar \\
I18308 & 20 & 18:33:33.05 & $-$08:39:18.0 & 1.33$\times$0.86 & 99 & 2.5 & 1.23 & 19.0$\pm$1.5 & 0.79$\pm$0.06 & no outflow & protostellar \\
I18308 & 21 & 18:33:34.22 & $-$08:38:32.0 & 0.77$\times$0.65 & 72 & 1.9 & 1.13 & - & - & no outflow & prestellar \\
I18308 & 22 & 18:33:31.56 & $-$08:39:01.0 & 0.96$\times$0.76 & 91 & 1.3 & 1.24 & 12.2$\pm$1.0 & 0.48$\pm$0.04 & outflow & protostellar \\
\enddata
\tablecomments{Table 4 is published in its entirety in the machine-readable format. Results for I18308 are shown here for guidance regarding its form and content.}
\end{deluxetable*}

\begin{figure*}[ht!]
\epsscale{1.0}\plotone{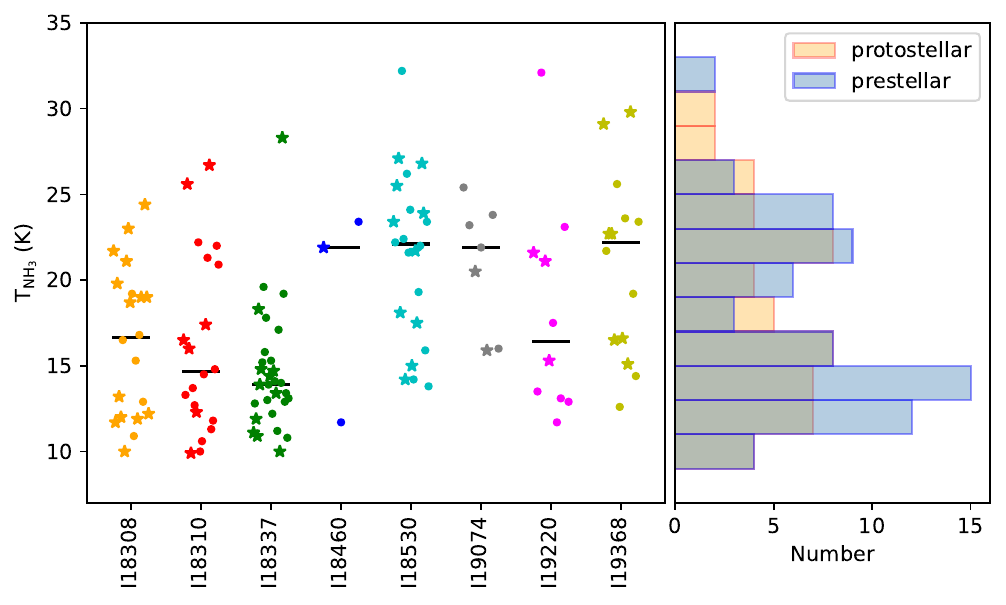}
\caption{{\it Left:} Gas temperatures of cores derived with \ammonia{} for all sources. The prestellar and protostellar cores are indicated with dots and stars, respectively. The black lines mark the median temperature for each source. {\it Right:} Distributions of core \ammonia{} temperatures for the prestellar and protostellar population.}\label{fig:Tgas}
\end{figure*}

We estimate gas temperatures of the cores with the VLA \ammonia{} (1,1) and (2,2) lines following the routine in \citet{Lu18}. A full account of the VLA observations will be published in other papers in the series. In summary, for each core we extract the average \ammonia{} (1,1) and (2,2) spectra within the FWHM returned by {\it getsf} and model the spectra assuming local thermodynamic equilibrium (LTE) conditions. We require a line detection $>$4~$\sigma$ for \ammonia{} (1,1) and $>$2.5~$\sigma$ for (2,2) for the fitting, and 125 cores out of 183 satisfy this criterion. The free parameters in the fit are the core velocity, rotational temperature, \ammonia{} column density, and velocity dispersion (see \autoref{sec:app_nh3}). During the fitting we assume a uniform filling factor of 1 for all transitions and hyperfine components. As illustrated in \autoref{fig:Tgas}, the measured $T_\mathrm{NH_3}$ ranges from 9.9 to 32.2~K, with a median of 16.5~K. The distribution for each region also varies. I18308, I18310, I18337, and I19220 have a lower median $T_\mathrm{NH_3}$ of 14--17~K, while I18460, I18530, I19074, and I19368 show a higher median $T_\mathrm{NH_3}$ of 22--23~K. 

The core mass is then estimated assuming that the emission purely comes from optically thin isothermal dust emission, enabling us to use the equation
\begin{equation}
    M_{\rm dust} = \frac{d^2F_\nu}{\kappa_\nu B_\nu (T_{\rm dust})}.
\end{equation}
where $d$ is the distance to the source, $F_\nu$ is the observed flux density, $B_\nu$ is the Planck function, $T_{\rm dust}$ is the dust temperature and $\kappa_\nu$ is the dust opacity at the observed frequency $\nu$. We adopt $\kappa_{\rm 1.3 mm}$ = 0.899~$\rm cm^2g^{-1}$ from \citet{Ossenkopf94} (thin ice mantles, $\rm 10^6~cm^{-3}$ density). We multiply the calculated dust mass by 100, assuming a dust-to-gas mass ratio of 1:100 \citep{Bohlin78}, to obtain the gas mass. In the absence of high resolution dust temperature measurements, we adopt the gas temperature derived with \ammonia. For those without valid \ammonia{} temperature measurements due to insufficient signal to noise ratio, we adopt the median \ammonia{} temperature for each region.

The determination of the evolutionary stages of dense cores, i.e., prestellar or protostellar, is essential to study their accretion history and interpret the observed mass distribution of core populations. This is, however, a difficult task in distant massive protoclusters. We first search for molecular outflows using CO~2--1, SiO~5-4 as well as SO, \htwoco{} and \methanol{} lines, and classify the associated dense cores as protostellar (Cheng et al. in prep.). Forty-three cores (43/183) are found to host evident molecular outflows. This is most likely a lower limit of protostellar cores since we will miss outflows orientated close to the plane of sky, and the outflow emission is often blended and confused in clustered environments. In light of this we follow \citet{Sanhueza19} and complement the identification of star forming cores by searching for associated line emission from ``warm core'' tracers, i.e., two \htwoco{} warm transitions ($3_{2,2}-2_{2,1}$) (218.475632~GHz, $E_u/k$=68.09~K), ($3_{2,1}-2_{2,0}$) (218.760066~GHz, $E_u/k$=68.11~K) and \methanol{} ($4_{2,2}-3_{1,2}$) (218.440063~GHz, $E_u/k$=45.56~K). Cores detected in these lines are likely to have been internally heated from star formation activities and hence they can be used as a star formation indicator. { In practice we examine both the morphology of the integrated line emission and the spectra for each core within the FWHM returned by {\it getsf}. If a peak above 4~$\sigma$ is found in any of the three spectral lines, and the emission has a centrally peaked compact morphology associated with dense cores, we define it as a warm core. This criterion gives 70 warm cores, of which 38 cores are also associated with outflows.} Cores that are absent in both molecular outflows and warm core tracers are then classified as prestellar candidates.  In summary, out of 183 cores in our sample, 108 are prestellar candidates and 75 cores are protostellar, including those identified by outflows (43/75) and warm core tracers (70/75). The fraction of prestellar cores significantly vary among the sample, with I18308 hosting a lowest fraction of prestellar population of 32\% (7/22) and I19220 a highest fraction of 76\% (13/17). 

\autoref{fig:Tgas} also reveals the difference in gas temperature of cores in the two populations. Overall the protostellar cores have a higher median $\rm T_{NH_3}$ of 17.5~K, compared to 15.8~K for the prestellar candidate cores. The $\rm T_{NH_3}$ distribution of both prestellar and protostellar cores appear to exhibit a bimodal shape, with one peak around 13~K and the other around 22~K, but the prestellar population has more cores in the lower temperature regime.



\section{Comparison of core identification algorithms: {\it getsf} v.s. {\it astrodendro}}\label{sec:core_iden_comp}

\begin{figure}[ht!]
\epsscale{1.0}\plotone{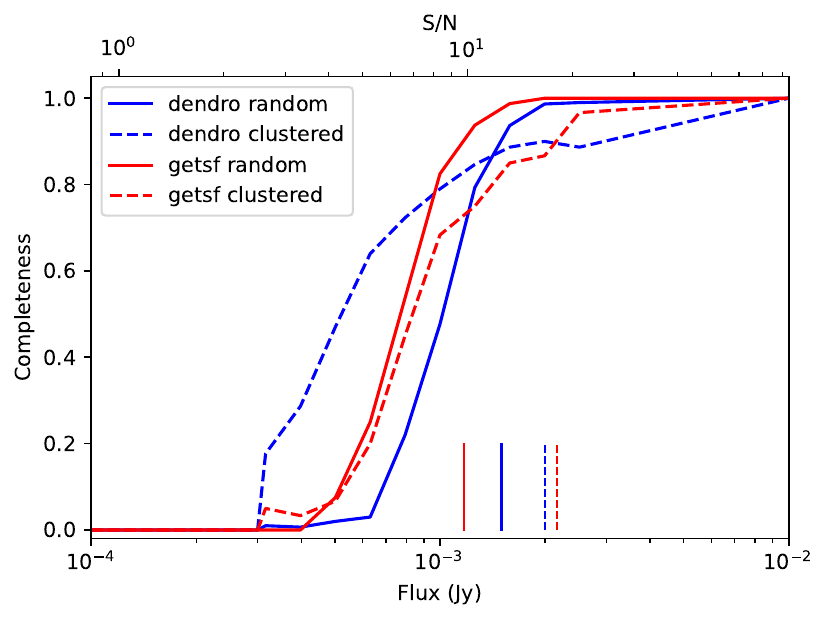}
\caption{Recovery fraction as a function of flux density for {\it astrodendro} and {\it getsf} (for I18308 as a representative). For artificial core insertion experiment two cases are explored: random insertion in the whole FOV of I18308 (primary beam correction $>$ 0.5), or random insertion only in regions where the flux density is greater than 1~\mjypbm (i.e., the relatively clustered region). The vertical lines denote the fluxes for which the detections are 90\% complete.}\label{fig:complete_comp} 
\end{figure} 

\begin{figure*}[ht!]
\epsscale{1.0}\plotone{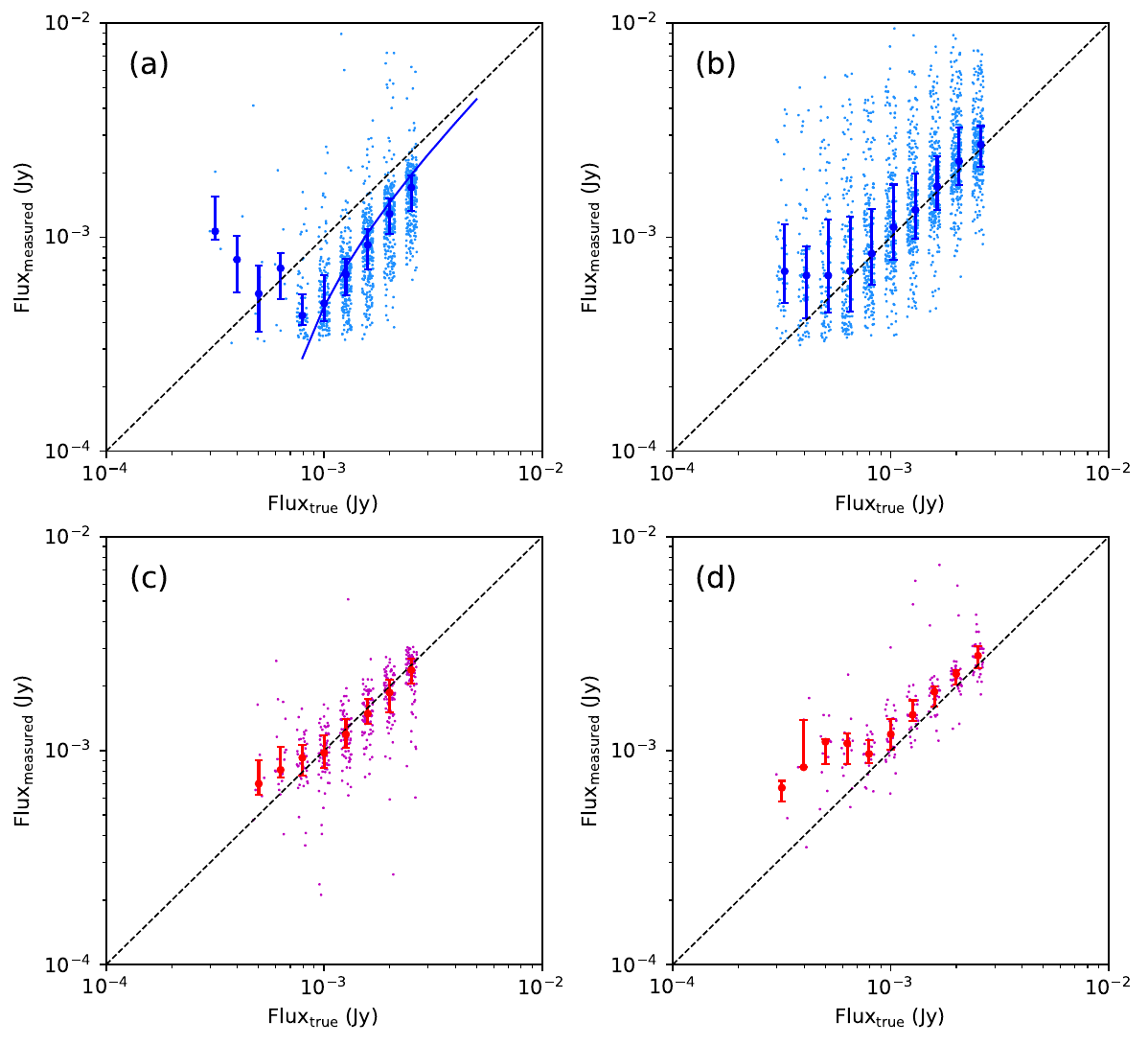}
\caption{ {\it (a)} Comparison between the true fluxes of the inserted cores and those returned by {\it astrodendro} (for the associated leaf structure). For each true flux value we manually shifted the data points in xaxis to better show the scattering level. Note that the errorbars here correspond to the first and third quartiles of the distribution for data points. The blue curve shows the expectation from \autoref{equ:reco}. {\it (b)} Same as {\it (a)} but the insertion is restricted in regions where the flux density is greater than 1~\mjypbm (i.e., the relatively clustered region). {\it (c)} Same as {\it (a)} but for {\it getsf}. {\it (d)} Same as {\it (b)} but for {\it getsf}. }\label{fig:complete_flux_comp}
\end{figure*}

Since the core identification results lay the foundation for further analysis of properties of the core population, it is necessary to review the algorithms adopted for core extraction. Cores, i.e., self-gravitating entities of gas and dust, usually manifest as local peaks in 2-D continuum or column density maps. There is a plethora of core extraction algorithms used in the literature, e.g., \textit{gaussclump} \citep{Stutzki90}, \textit{clumpfind} \citep{Williams94}, \textit{astrodendro} \citep{Rosolowsky08},  \textit{blobcat} \citep{Hales12}, \textit{fellwalker} \citep{Berry15}, and \textit{getsf} \citep{Men21}, etc. A complete assessment of the performance of all these algorithms is clearly beyond the scope of this paper. In this section we compare the performance of {\it getsf} and {\it astrodendro} in detecting and extracting properties of cores. Both are very widely used in the literature: {\it astrodendro} is designed for identifying hierarchical structures in continuum or line data \citep[e.g.,][]{Ginsburg16,Sanhueza19,ONeill21,Takemura22}. {\it getsf} (or its predecessor algorithms {\it getsources}, {\it getfilaments}, and {\it getimages}, \citet{Menshchikov12,Menshchikov13,Menshchikov17}), is the fiducial method in this paper and also commonly used in other works for multi-scale multi-wavelength source-extraction, including the ALMA-IMF large program survey \citep[][]{Motte22,Pouteau22}. Therefore, when putting in context the statistical analysis of the INFANT core population, e.g., the core mass function, it is important to understand the systematic differences that may arise from different methodologies.

In \autoref{fig:map_iden} we show the core identification results for I18308 and I18337 using both {\it dendrogram} and {\it getsf} methods. The same plot for other regions are shown in \autoref{sec:app_comp_method}. For {\it getsf} the setup is described in \autoref{sec:core_iden}. For {\it dendrogram} we set the base flux density threshold to 4$\sigma$, the minimum significance for structures to 1$\sigma$, and the minimum area to half the synthesized beam. This choice of parameter setup has been used and works reasonably well in a series of CMF works \citep[][]{Cheng18,Liu18,ONeill21}. 
Inspection of the image allows one to assess how different algorithms operate on the imaging data. {\it astrodendro} appears to identify more cores compared to {\it getsf}. While most bright, high contrast cores are identified in both methods, {\it astrodendro} also includes some relatively weak cores that often exhibit irregular boundaries, which are not picked up by {\it getsf}. In fact for all the eight clouds {\it astrodendro} returns a total of 382 cores, about twice the core number by {\it getsf}. 
Another feature is that for the brightest and most massive cores {\it getsf} tends to return more fragments.

To quantitatively evaluate the performance of the two algorithms, we run experiments of artificial core insertion. This is the same as done in \autoref{sec:core_iden} for {\it getsf} but here we check the detection rate and also recovery of core fluxes for both algorithms. The number recovery is shown in \autoref{fig:complete_comp}. If the artificial cores are randomly inserted in the original image (i.e., 50\% primary beam response, the same criterion with which we identify cores), the 90\% completeness limit is estimated to be 1.5~mJy (or S/N = 12.5) for {\it astrodendro}, and 1.2~mJy (S/N = 9.8) for {\it getsf}. Therefore, {\it getsf} is slightly more efficient than {\it astrodendro} in detecting weak sources.

We also test the case with artificial cores restricted to the spatial range defined by a flux density threshold, e.g., $\sim$1~\mjypbm{} level, which means they preferably show up in crowded environments. Blending and confusion make it more challenging to pick up cores. The 90\% completeness limit increases to be 2.1~mJy (S/N = 17.7) for {\it astrodendro} and 1.9~mJy (S/N = 15.5) for {\it getsf}. It is worth noting that for {\it astrodendro} the detection completeness for low S/N ($\sim$2--6) cores is significantly higher than that in the random insertion case. By definition a local peak has to be greater than 5$\sigma$ in order to be identified by {\it astrodendro} with our parameter setup. The enhanced detection fraction for low S/N (2--6) cores reflects the fact that these cores lie on strong background emission in clustered regions, thus having a larger ($>5\sigma$) peak value, and {\it astrodendro} does not distinguish the large scale emission from the locally condensed structures. This also partly explains why {\it astrodendro} returns much more cores than {\it getsf} while having a larger 90\% completeness limit flux in the random insertion case. We do not take this as a merit for {\it astrodendro} (in terms of core identification) since the detection of cores in isolation or crowded fields (with strong large scale emission) may essentially have been treated with different significance threshold, and the false detection rate could also be elevated when background emission is strong. For example, artificial fragments/substructures with low or moderate S/N could be produced when a smooth flux distribution is observed with interferometers \citep[e.g.,][]{Hodge16,Caselli19}. 

In \autoref{fig:complete_flux_comp} we show how well the core fluxes are recovered with two methods. {\it astrodendro} is an algorithm based on pixel assignment, and the flux of a leaf is the sum of fluxes in all leaf pixels in an isophotal boundary defined by input base value (in case of an isolated leaf). When this flux is directly interpreted as the core flux, there will be two factors that lead to deviation from the true value. First, pixels below the base level (4~$\sigma$) are not assigned to any core structures, rendering the core flux underestimated. This underestimation fraction is greater for lower flux cores, which is analytically tractable for point sources with peak flux density $f_{\rm peak}$,
\begin{equation}\label{equ:reco}
\frac{F_{\rm measure}}{F_{\rm true}} = 1-\frac{f_{\rm base}}{f_{\rm peak}},
\end{equation}
where $f_{\rm base} = 4\sigma$ is our case. This type of flux underestimation is dominating in the random insertion case for {\it astrodendro}, as illustrated in panel (a) of \autoref{fig:complete_flux_comp}. On the other hand, the pixels assigned to a leaf structure may well contain fluxes contributed by larger scale emission, especially in clustered regions. As seen in panel (b), this effect will result in a systematically higher measured flux and also a larger scattering since the measured values are dependent on emission level of the background, which often varies in different locations. In comparison, the fluxes returned by {\it getsf} are in general more robust and show less scattering.

In summary, {\it getsf} has a better treatment of the varying background levels in identification of dense cores, and gives more robust flux measurements. Cautions should be made when directly interpreting the leaf structures given by {\it astrodendro} as cores. Note that the comparison here only concerns the performance of algorithms in detecting point-like sources in a 2-D map. In principle, it is likely that these 2-D cores do not always correspond to real physical dense cores when taking account the  effects of projection and other observational complications (radiative transfer effects, interferometric artifacts, instrumental noise, etc.) \citep[e.g.,][]{Padoan23}.
Hereafter the discussion on the INFANT core population is based on the core catalog with the {\it getsf} method, but we also discuss in \autoref{sec:app_comp_method} about the potential influence on the statistical results if {\it astrodendro} is used, such as the core mass function (CMF).

\section{Discussion}\label{sec:discussion}

\subsection{Evolutionary status of the protocluster}\label{sec:evolution}

Complete census of dense cores in protoclusters requires high sensitivity and high spatial resolution observations covering a large area ($\gtrsim$1~pc$^2$) and being able to resolve the cloud down to a few 1000~au scale at the same time. Even in the ALMA era, large sample surveys of protoclusters that could achieve statistically significant conclusion are still challenging. Surveys to date include ASHES \citep{Sanhueza19}, ALMA-IRDC \citep{Barnes21}, and ALMA-IMF \citep{Motte22}, etc. 

Different from surveys that focuses on the very early evolutionary stages in infrared dark clouds, our INFANT survey covers a range of evolutionary phases. Overall our targets are more evolved compared to the 70~$\mu$m dark clouds in \citet{Sanhueza19} given the overall higher luminosity and the fact that all the targets already host bright point sources in the near- and mid infrared (see \autoref{sec:app_sed}), indicating ongoing protostellar activities. But still a significant portion of the cloud can be dark at wavelengths up to 24~$\mu$m, i.e., for I18308, I18310, I18337, I18460, or 8~$\mu$m for I18530 and I19220, where the star formation has not yet begun or is still in an early phase. This is expected as a parcsec scale cloud may well contain star formation sites in various stages. In I18530, 19074 and I19220, compact 1.3~cm continuum emission is detected (Cheng et al. in prep.), suggesting the protostars have evolved into the ultracompact (UC) H II region phase. An even more evolved phase is found in I19074, where a parsec-scale H II region traced by an infrared bubble is detected (\citealt{Churchwell06}, see also \citealt{Lu18}), thus the observed dust/gas content could be the adjacent material compressed by the HII region and/or the remnant after the cloud complex has been destructed by feedback.

In \autoref{table:info_sample} we have listed the luminosity-to-mass ratio $L/M$, which is often used as a proxy for the evolution of parsec scale clumps \citep[e.g.,][]{Ma13,Liu16,Molinari16}. More evolved sources in general produce more infrared emission and have consumed more gas material, thus leading to higher luminosity-to-mass ratios. The mass and luminosity are derived with the temperature and column density map from {\it Herschel} SED fitting (see \autoref{sec:app_sed}), and the measurement is made for the spatial range used for ALMA imaging. The $L/M$ ratios range from 2.1 $L_\odot/M_\odot$ (I19368) to 12.7 $L_\odot/M_\odot$ (I19220). According to this ratio, the sample can be roughly divided into two groups, i.e., I19368, I18337, I18460, I18308 that are in a relatively early stage, while I18310, I18530, I19220 and I19074 that are relatively evolved. This is broadly consistent with other evolution indicators such as infrared emission, and UCH~II/HII regions. 

Two exceptions are I19368 and I18310. The I19368 region has the lowest $L/M$ ratio, but widespread protostellar activity is seen in the infrared images. I18310 is included in the more evolved group, but part of the cloud is still infrared dark for wavelengths up to 24~$\mu$m. This may reflect the limitation of $L/M$ as an evolution indicator for a parsec-scale cloud, especially when the difference in $L/M$ is not significant. For example, a newly formed high mass young stellar object could rapidly boost the observed $L/M$ considering a typical luminosity–mass relation for main sequence stars \citep[e.g.,][]{Eker15}. Therefore, the relatively low $L/M$ of I19368 compared to other regions, but widespread infrared emission, may arise from a lack of high mass young stars in this region. Since the $L/M$ indicator alone is subject to the inherent stochasticity of the sampling of protostellar masses across the star forming cloud, we only consider sources with consistent evolution classifications using either $L/M$ and other evolution indicators for following discussions, i.e., the young group includes I18308, I18337 and I18460; and the evolved group includes I18530, I19074 and I19220.


\subsection{Absence of high-mass prestellar core and mass growth of cores}\label{sec:core_growth}


\begin{deluxetable*}{lcccc}[!t]
\tabletypesize{\scriptsize}
\tablecaption{Mann-Whitney~U~test for comparison between core populations \tablenotemark{a}\label{table:mass_comp}}
\tablewidth{0pt}
\tablehead{
\colhead{{Sample}}  & {1.3~mm Flux density} & & {Mass} 
}
\startdata
(All)~~~~~~~~~prestellar v.s. protostellar  & 1.5$\times10^{-8}$ &    & 4.0$\times10^{-7}$ \\
(Young)~~~ prestellar v.s. protostellar    & 5.0$\times10^{-5}$ &   & 1.4$\times10^{-4}$ \\
(Evolved)\ prestellar v.s. protostellar    & 3.6$\times10^{-3}$ &   & 4.7$\times10^{-3}$ \\
\enddata
\tablenotetext{a}{$p$ values of the Mann-Whitney~U~test \citep{Mann47} for the samples for a single-sided (less) test, i.e., the null hypothesis is that the protostellar flux/mass is equal or smaller than that of the prestellar population.}
\end{deluxetable*}

\begin{figure*}[ht!]
\epsscale{1.1}\plotone{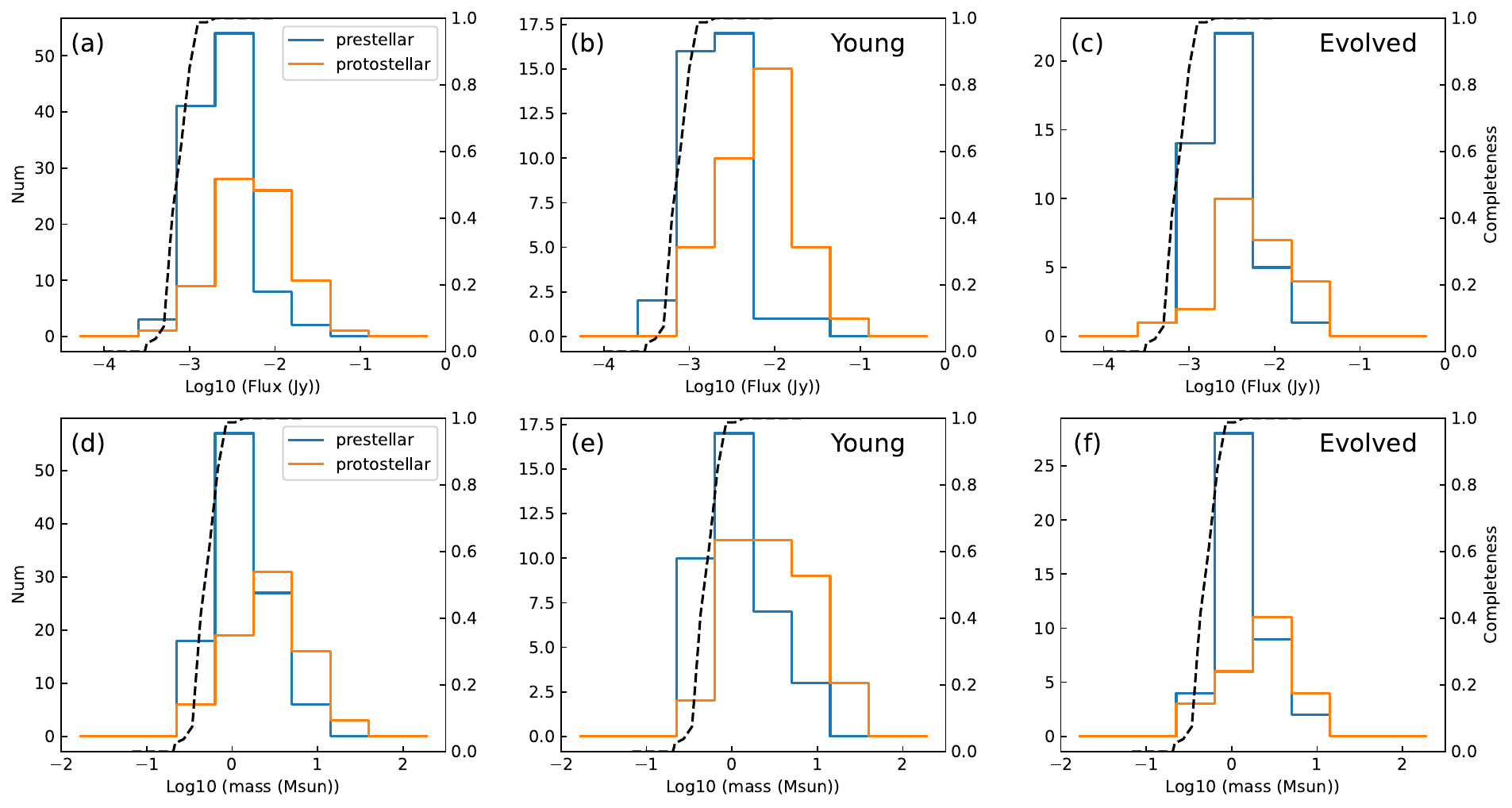}
\caption{{\it (a)} 1.3~mm flux distribution for the prestellar candidate cores and protostellar cores. The dashed black line indicates the recovery fraction (or completeness level) estimated for I18310 as a representative. {\it (b)} Same as {\it (a)} but only including cores from the relatively early type clouds, defined in \autoref{sec:evolution}, i.e., I18308, I18337, I18460. {\it c)} Same as {\it (a)} but only including cores from the relatively late type clouds, defined in \autoref{sec:evolution}, i.e., I18530, I19220, I19368. {\it (d)} Mass distribution for the prestellar candidate cores and protostellar cores, where the masses are estimated using the \ammonia{} gas temperature. The dashed lines indicate the estimated recovery fraction for I18310 as a representative. A temperature of 14~K and 19~K, i.e., the median \ammonia{} temperature for the two populations, are adopted for the red and blue line, respectively. {\it (e)} Same as {\it (d)} but only including cores from the relatively early type clouds, i.e., I18308, I18337, I18460. {\it (f)} Same as {\it (d)} but only including cores from the relatively early type clouds, i.e., I18530, I19220, I19368.
}\label{fig:comp_mass}
\end{figure*}

{ Theories to explain high mass star formation can be broadly categorized into ``core-fed'' and ``clump-fed'', each predicting distinct initial conditions and spatial scales for the mass reservoir that serves as the source for accretion \citep{Wang10}. One way to test different formation theories is to characterize the initial conditions, and in particular, to examine whether there exist high mass, gravitationally bound, pre-stellar cores as predicted by the core accretion theories \citep{Mckee03}. A handful of candidates have been identified \citep[e.g.,][]{Cyganowski14,Kong18,Barnes23}, { but none of them has been confirmed as bona-fide high-mass prestellar cores.} In this survey the most massive prestellar core, namely c2 in I18337, has a mass of only $\sim$10.3~\msun, and in other regions the most massive prestellar core are all below 10~\msun. Even when adopting a high core-to-star efficiency of 50\% \citep[e.g.,][]{Tanaka17}, and assuming no multiplicity, a core with an initial mass of approximately 16~\msun{} is necessary to give rise to an 8~\msun{} star. Thus, there are no high-mass prestellar cores in this survey. This negative outcome is similar to those found in other surveys \citep[e.g.,][]{Sanhueza19,Mori23}, suggesting that high-mass prestellar cores, if they do exist, are very rare.}


{ A possible scenario to form massive stars without invoking a high-mass prestellar core entails substantial growth in core mass after their initial formation.} To investigate this scenario we compare the mass distribution of the prestellar and prostellar core populations. \autoref{fig:comp_mass}(a) shows the 1.3~mm flux distribution of two populations combining cores from all clouds. The protostellar cores tend to have larger 1.3~mm fluxes, with a median of 5.51~mJy, compared to 2.56~mJy of the prestellar cores. We further plot in \autoref{fig:comp_mass}(d) the distributions of masses  derived assuming the \ammonia{} temperatures. {The median mass of protostellar cores (2.81~\msun) is also larger than that of the prestellar population (1.34~\msun).}
We carry out a Mann-Whitney U test \citep{Mann47} to statistically examine the result, as shown in \autoref{table:mass_comp}. It is a nonparametric test for the ranking between two samples. For panel (a) the null hypothesis that protostellar cores have equally or smaller fluxes than prestellar cores can be rejected with a confidence greater than 99.99\% ($p$ = 1.5$\times 10^{-8}$). Similarly for panel (d) the null hypothesis that protostellar cores have equal or smaller masses than prestellar cores can also be rejected with a high confidence ($p$ = 4.0$\times 10^{-7}$). Therefore the protostellar cores are statistically more massive. We also show in \autoref{sec:app_virial} that the same conclusion also applies for the gravitationally bound core population with virial parameters smaller than 2, where the virial parameters are estimated using the \ammonia{} line width.  

Note that the core sample is flux limited and a temperature is needed to convert the flux to mass. Thus our observations are more sensitive for protostellar cores compared to prestellar cores, since the former class has on average a higher temperature, thus a lower mass limit accordingly. For example, a 90\% completeness flux of 1.2~mJy (typical value for our sample) translates into a mass of 0.55~\msun{} assuming 20~K, but 1.5~\msun{} assuming 10~K, i.e., a factor of 2.7 larger. Therefore, it is likely that we are missing more prestellar cores in the lower mass regime.

{ We further check sub samples of clouds in different evolutionary stages, as shown in \autoref{fig:comp_mass}(e) and (f). For the young group (I18308, I18337, I18460), the protostellar cores are clearly more massive than prestellar cores (3.29~\msun{} v.s. 1.16~\msun{} in median mass). The null hypothesis that protostellar cores have equal or smaller masses can be rejected with a confidence greater than 99.9\% ($p$=1.4$\times 10^{-4}$).  On the other hand, for the more evolved group (I18530, I19220, I19074), the difference between two populations is smaller (2.31~\msun{} v.s. 1.37~\msun{} in median mass), but still a $p$ value of 4.7$\times 10^{-3}$ suggests the protostellar population is more massive than the prestellar population with a confidence greater than 99\%. This trend would be reinforced when taking into account the different completeness levels for the two populations.}



Similar to our results, larger masses for protostellar cores have also been observed in other star forming regions, i.e., IRDCs, Orion~A, and W43 \citep{Peretto20,Kong21,Takemura23,Nony23,Li23}, and interpreted as evidence for continuous mass accretion during the protostellar phase. Such mass growth is expected in the clump-fed models of star formation \citep{Smith09}, such as competitive accretion \citep[e.g.,][]{Bonnell06}, global hierarchical collapse model \citep[]{Vazquez19}, or the inertial-inflow model \citep{Padoan20}, and provides a pathway to form massive stars without previously existing massive prestellar cores. The accretion onto cores could be filament-mediated as shown in some case studies, i.e., the gas flow along filaments feeds the embedded cores and protostars with additional mass \citep[e.g.,][]{Lu18}. In our case, protostellar cores have a median mass about twice as large as prestellar cores. To make two distributions indistinguishable, e.g., $p$ $\sim$0.1 in a single-side Mann-Whitnery U test, the prestellar masses need to be multiplied by a factor greater than 3.2 and 1.8 for the young and evolved group, respectively. In a core growth scenario, this suggests a core has to gain a significant fraction of its total mass via further accretion during its lifetime. If this happens in the protostellar phase, the rate at which a core accretes from its surroundings must be on average larger than the protostellar accretion rate onto the accretion disk. 

\subsection{Core mass function}\label{sec:cmf}

\begin{figure*}[ht!]
\epsscale{1.1}\plotone{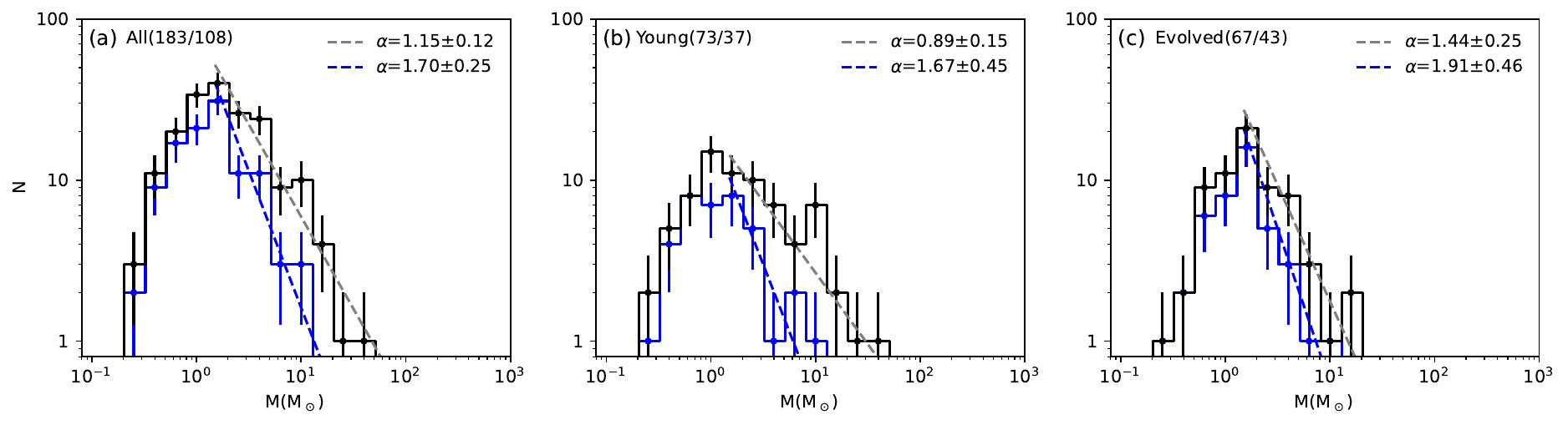}
\caption{{\it (a)} CMF and prestellar CMF of the INFANT sample shown in black and blue, respectively. The core masses are calculated using the \ammonia{} gas temperatures. The dashed lines show the best power-law fit result for the high-mass end (M $>$ 1.5~\msun) for the corresponding CMFs. {\it (b)} Same as {\it (a)} but only including cores from the relatively early type clouds, i.e., I18308, I18337, I18460.  {\it (c)} Same as {\it (a)} but but only including cores from the relatively late type clouds, i.e., I18530, I19220, I19368. }\label{fig:cmf}
\end{figure*} 

The distribution of stellar masses in young clusters, i.e., the initial mass function (IMF), is a fundamental outcome of star formation. The shape of the IMF and whether it is universal are important topics in modern astrophysics. Observations in nearby clusters have revealed remarkably universal IMFs \citep{Bastian10}, which appear to follow a power law index above 1~\msun,
\begin{equation}
\frac{dN}{d\log M} \propto M^{-\alpha},
\end{equation}\label{equ:imf}
\noindent 
where $\alpha=$1.35 \citep{Salpeter55}, with a peak near 0.25~\msun. The physical origin of the IMF is unclear. A promising way to advance our understanding is to look at the dense core mass function, where a core can be defined theoretically to be a self-gravitating structure that collapses to form a single star or small multiple system. The CMF in nearby, $\lesssim$1 kpc regions is similar in shape to the IMF \citep[e.g.,][]{Motte98,Alves07}. Studies with ALMA have shed more light on the CMF in more distant regions \citep{Zhang15, Csengeri17,Cheng18,Liu18,Motte18,Kong19,Sanhueza19,Lu20,Sadaghiani20,ONeill21,Suarez21,Pouteau22,Pouteau23}. Some of these measurements, which compile a sample of single pointing observations towards massive clumps \citep[e.g.,][]{Liu18,ONeill21}, are likely biased due to potential mass segregation effect \citep{Plunkett18,Dib19}, thus it is necessary to map the protocluster over their full extent. In this section we examine the CMF of the INFANT sample. 

In \autoref{fig:cmf}(a) we show the CMF of 183 cores identified by {\it getsf}. The CMF of 108 prestellar cores is shown in blue. 
As shown in \autoref{sec:core_iden}, the 90\% completeness limit is estimated to be 0.53--0.89~\msun{} (for $T=$15~K), which is lower than the mass range where CMF starts to peak or flatten, i.e., 1 -- 3~\msun. But considering cores with a lower temperature of 10~K, the  90\% completeness limit can be higher, i.e., 1.0 -- 1.5~\msun, thus we can not say with certainty whether the turnover seen in the CMF is physically real or due to incompleteness. We fit the high-mass end of the CMFs (M $>$ 1.5~\msun) with a power-law function as in \autoref{equ:imf} using the the maximum likelihood estimation (MLE) method in \citet{Clauset09} implemented with the {\it plfit} package. The results are labeled in \autoref{fig:cmf}.
The fit gives an power law index $\alpha$ = 1.15 $\pm$ 0.12. This slope is similar to that found in the IRDC sample in \citet{Sanhueza19} (1.07 $\pm$ 0.09 for all cores, or 1.17 $\pm$ 0.10 for prestellar cores), and shallower than the Salpeter value of 1.35. For the (candidate) prestellar core population, the power law index is measured to be $\alpha = 1.70 \pm 0.25$, slightly steeper than the Salpeter value. Thus the slightly top-heavy form measured for the global CMF is due to the protostellar population. Similar behavior is also observed in the W43 cloud, where the global CMF has a high-mass slope significantly shallower than the prestellar CMF (0.96$\pm$0.09 v.s. 1.46$_{-0.19}^{+0.12}$) \citep{Nony23}. This flattening trend could be explained by core mass growth in clump-fed models if high-mass cores accrete more efficiently than low-mass cores \citep{Nony23}.

To investigate the possible dependence on the overall cloud evolutionary status we plot the CMF for the young and evolved cloud groups, respectively, in \autoref{fig:cmf} (b) and (c). 
For the young group, we get a top heavy power-law index $\alpha$ = 0.89 $\pm$ 0.15, while for the evolved group, the slope is steeper $\alpha$ = 1.44 $\pm$ 0.25, consistent with the Salpeter value. Thus we see a trend of the global CMF becoming steeper at the high mass end as the protocluster evolves, resembling more the IMF shape at later stages. We also fit the prestellar CMF for these sub samples, the slopes are steeper compared to the global CMF but less well constrained due to limited number statistics in the high mass end. 


This steepening trend with cloud evolution is seemingly in contradiction to the speculation that protostellar core growth could flatten the CMF. Combined with our analysis of core growth in \autoref{sec:core_growth}, this evolutionary variation in CMF could be explained in a dynamic cluster formation picture. If the prestellar cores, which constantly form over time and evolve into the protostellar population, all share the same mass distribution as the ``parent'' prestellar core population, then in order to make a shallower slope, high-mass cores need to have a larger fractional mass growth rate than low-mass cores. Precisely, the net mass accretion rate $\dot{M}$ should scale with mass with an power-law index significantly larger than 1. This agrees with observations in some hub-filament systems, where global infall results in continuous gas inflow from the clump scale to the core scale and efficiently feeds the most massive core(s) early in the evolution, leading to a high mass concentration fraction of the massive cores \citep[e.g.,][]{Anderson21,Zhou22,Xu23}. Therefore, in the very early stage, which is not captured by our sample, the CMF originated from fragmentation could be Salpeter-like, or slightly top-heavy as suggested in \citet{Sanhueza19}, but rapid accretion from the clump then produces a shallower global CMF, as we see for the young group. In a later stage, the clump-fed accretion will be reduced as protostellar feedback becomes stronger, rendering it difficult to continue forming more massive cores. The ongoing formation of low- and intermediate mass cores leads to a steeper, Salpeter-like slope as we observe for the evolved group. 

To confirm or refute this interpretation, it is necessary to conduct a systematic study of the core accretion rate for different core masses and different cloud evolutionary stages. We intend to further explore this using our ALMA band 3 data. As previously mentioned there are numerous measurements of power-law indices of the CMF in massive clumps, with a range of CMF slopes at high-mass end ($>$1~\msun) from a top heavy value of $\sim$0.7 to a Salpeter-like one of 1.35. Here we do not attempt to reconcile these measurements in the literature, since the core identification and mass estimation  methods vary. In order to make a meaningful comparison with our work we consider studies that (i) image the protocluster in its full extent; (ii) include temperature measurements at the core scale; and (iii) adopt a similar core detection algorithm as {\it getsf}. There are very a few works that satisfy these criteria. \citet{Motte18} found $\alpha$ = $-$0.90 $\pm$ 0.06 for masses M $>$ 1.6$\msun$ from a sample of 105 cores in the massive cloud W43-MM1. More recent results from the ALMA-IMF survey also revealed a top heavy index $\alpha$ = $-$0.95 $\pm$ 0.04 for in a sample of 294 cores in W43-MM2\&MM3 \citep{Pouteau22}. In the scenario proposed here, these measurements in W43 could represent the relatively early, active accreting phase in the cluster evolution.  

A caveat in this survey is concerning the classification of two groups with different evolutionary phases. As discussed in \autoref{sec:evolution}, It is difficult to unambiguously define and determine the relative evolutionary phase for a parsec-scale protocluster. In addition, since each group contains only three clouds, it is likely that the observed differences in CMF may (partly) arise from other systematic differences of the protoclusters, instead of evolution. For example, the targets in the young group also have relatively higher masses. A larger sample with similar spatial resolution is needed to disentangle the coupling.

\section{Conclusions}\label{sec:summary}
We present the first results of the INFANT (INvestigations of massive Filaments ANd sTar formation) survey, aimed at investigating the relationship between filaments and star formation across a range of evolutionary stages. In this first paper of the series, we report the ALMA Band~6 continuum observations of eight carefully selected massive filamentary clouds that have been characterized in previous SMA and VLA observations \citep{Lu14,Lu18}. Each cloud was mosaiced in 30 to 40 pointings in 12~m array (5000 -- 6000~arcsec$^2$), with a sensitivity of $\sim$0.1~\mjypbm{} and a spatial resolution of $\sim$3000~au. The main findings are summarized as follows: 

\begin{itemize}
\item[1.] 
We compare the performance of two widely used core identification algorithms, {\it getsf} and {\it astrodendro}, via artificial core insertion experiments. We find that {\it getsf} gives more robust identification results for identification of compact cores and flux measurements in both isolated and clustered region. Cautions should be taken when directly interpreting the leaf structures given by astrodendro as cores. 

\item[2.]
We detect in total 183 dense cores for the INFANT sample, with each cloud having 15 -- 36 cores. Each core is classified as either protostellar if it is associated with molecular outflows and/or warm gas tracers including \htwoco{} ($3_{2,2}-2_{2,1}$), ($3_{2,1}-2_{2,0}$), \methanol{} ($4_{2,2}-3_{1,2}$), or prestellar candidates if not. Out of 183 cores 108 are classified as prestellar and 75 are prototellar, including those identified by outflows (43/75) and warm core tracers (70/75). The fraction of prestellar cores varys among the sample, with I18308 hosting a lowest fraction of prestellar population of 32\% (7/22) and I19220 a highest fraction of 76\% (13/17).

\item[3.]
Our INFANT targets cover different cluster evolutionary stages based on presence of infrared emission and UCHII/HII regions. We divided the INFANT sample into two groups with different evolutionary stages, young group (I18337, I18460, I18308) and evolved group (I18530, I19220, I19074) based on the $L/M$ ratio, as well as other evolution indicators. Two targets, I19368 and I18310, are not included since we cannot assign them an evolution group confidently.

\item[4.]
We estimate the gas temperature of cores using the \ammonia{}(1,1) and (2,2) lines obtained with JVLA following a spectral line fitting routine in \citet{Lu18}. The measured $T_\mathrm{NH_3}$ ranges from 9.9 to 32.2~K, with a median of 16.5~K. The protostellar cores have a higher median $\rm T_{NH_3}$ of 17.5~K, compared to 15.8~K for the prestellar candidate cores. 

\item[5.]
With masses derived with the 1.3 mm continuum emission and \ammonia{} temperatures, we compare the mass distribution between protostellar and prestellar populations. We find that the protostellar cores are statistically more massive than the prestellar cores. This provides evidence that cores continue to gain mass from the surroundings after the formation of protostars.

\item[6.]
The most massive prestellar core in this sample has a mass of 10.3~\msun. We have not detected high-mass ($\gtrsim$20--30~\msun) prestellar cores.

\item[7.]
For the INFANT sample we derive a powerlaw index of 1.15 $\pm$ 0.12 for the global CMF, and 1.70 $\pm$ 0.25 for the prestellar CMF in the high mass end (M $>$ 1.5~\msun). We also find a trend for steeper power-law index in global CMF with cloud evolution (1.44 $\pm$ 0.25 for the evolved group v.s. 0.89 $\pm$ 0.15 for the young group). An explanation is that cores could continuously grow by acquiring infalling material from larger scale after their formation. This process is particularly strong in the relative early stage of cluster formation, leading to a shallower CMF as more massive cores accrete more efficiently. At a later point when core accretion is reduced due to protostellar feedback, it is more difficult to continue forming more massive cores, thus ongoing formation of low- and intermediate mass cores results in a steeper, Salpeter-like slope.

\end{itemize}

{Identifying the core population is the first step towards understanding the importance of filaments in high-mass star formation. In the subsequent papers of this series, we will utilize molecular line data to delve into the significance of filamentary morphology. This will involve examining phenomena such as gas flows and accretion along the filaments, as well as comparing gas kinematics in star-forming filaments with those in more quiescent filaments.}

\acknowledgments
This work has been supported by the National Key R\&D Program of China (No. 2022YFA1603101). This work has been sponsored by Natural Science Foundation of Shanghai (No. 23ZR1482100). X.L. acknowledges support from the National Natural Science Foundation of China (NSFC) through grant No. 12273090, and the Chinese Academy of Sciences (CAS) ‘Light of West China’ Program (No. xbzg-zdsys-202212). This paper makes use of the following ALMA data: ADS/JAO.ALMA\#2017.1.00526.S. ALMA is a partnership of ESO (representing its member states), NSF (USA) and NINS (Japan), together with NRC (Canada), MOST and ASIAA (Taiwan), and KASI (Republic of Korea), in cooperation with the Republic of Chile. The Joint ALMA Observatory is operated by ESO, AUI/NRAO and NAOJ. The National Radio Astronomy Observatory is a facility of the National Science Foundation operated under cooperative agreement by Associated Universities, Inc. R.G.-M. acknowledges support from UNAM-PAPIIT project IN108822. H. B. L. acknowledges the support from the National Science and Technology Council (NSTC) of Taiwan (Grant No. 111-2112-M-110-022-MY3). P. S. was partially supported by a Grant-in-Aid for Scientific Research (KAKENHI Number 22H01271) of JSPS.

\vspace{5mm}

\software{CASA \citep{Mcmullin07,casa22}, APLpy \citep{Aplpy12}, Astropy \citep{Astropy13,astropy18,astropy22}}

\clearpage
\appendix
\counterwithin{figure}{section}
\counterwithin{table}{section}

\section{Multi-wavelength images of the INFANT sample}\label{sec:app_sed}
We present in \autoref{fig:multi1} the infrared images of the INFANT targets including {\it Spitzer} 3.6/8.0 $\mu$m, WISE 12/22 $\mu$m as well as the column density/dust temperature maps derived with {\it Herschel} data (70$\mu$m, 160$\mu$m from PACS and 250$\mu$m, 350$\mu$m, 500$\mu$m from SPIRE). 
For the SED fits we assume a single-component, modified black-body spectral energy distribution fits to each pixel of input images.
Before performing SED fitting, we smoothed all images to a common angular resolution of the largest telescope beam ($\sim$37\arcsec for SPIRE 500$\mu$m image) and all images were re-gridded to have the same pixel size.
The flux density is given by
\begin{equation}
S_\nu = \Omega_mB_\nu(T_\mathrm{d})(1-e^{-\tau_\nu}),
\end{equation}
and the column density is
\begin{equation}
N_\mathrm{H_2} = \frac{\tau_\nu M_g}{\kappa_\nu \mu m_\mathrm{H}M_d},
\end{equation}
where $B(T_d)$ is the Planck function at $T_d$ and $\kappa$=$\kappa_{230}(\nu/230GHz)^\beta$, $\Omega_{m}$ is the solid angle. We adopt $\kappa_{230}$ = 0.899 \gcm{} as in \citet{Ossenkopf94} for the moderately coagulated thin ice mentle model and a gas to dust ratio $M_g/M_d$ = 100 \citep{Bohlin78}. 
We performed least-squares fits of the 70/160/250/350/500$\mu$m spectral energy distributions weighted by the squares of the measured noise levels to derive the pixel-to-pixel distributions of the column density/dust temperature maps with an angular resolution of $\sim$37\arcsec.

\begin{figure*}[ht!]
\epsscale{1.0}\plotone{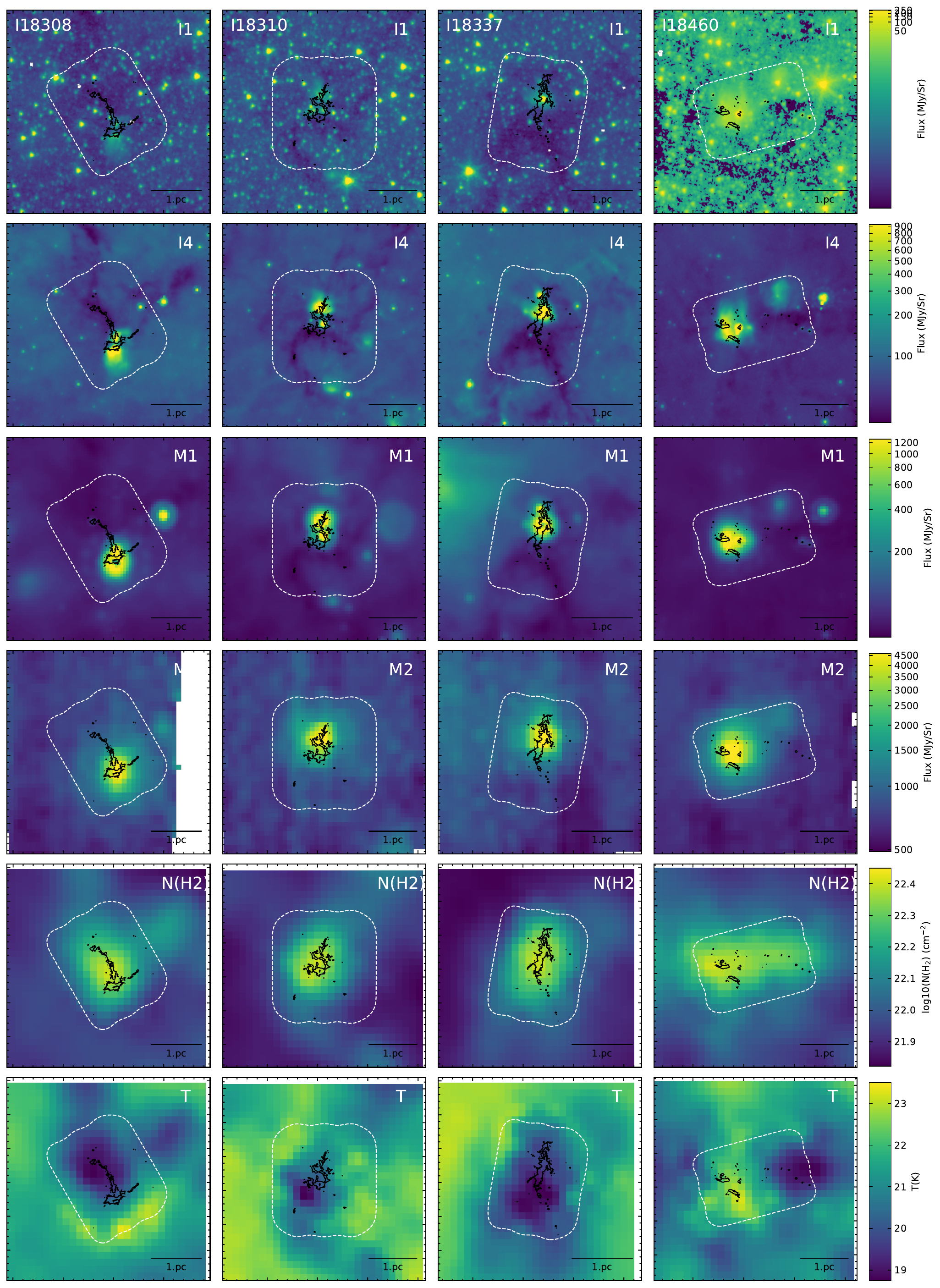}
\caption{Multi wavelength images of the INFANT targets (from left to right: I18308, I18310, I18337 and I18460). From top to bottom we present the images of Spitzer 3.6/8.0 $\mu$m, WISE 12/22 $\mu$m, and column density/dust temperature maps constructed with the {\it Herschel} data (see text). The white boxes indicate regions mapped by ALMA in band~6. The black contour shows the ALMA 1.3~mm at 5~$\sigma$ level.}\label{fig:multi1}
\end{figure*} 
\addtocounter{figure}{-1}
\begin{figure*}[ht!]
\epsscale{1.0}\plotone{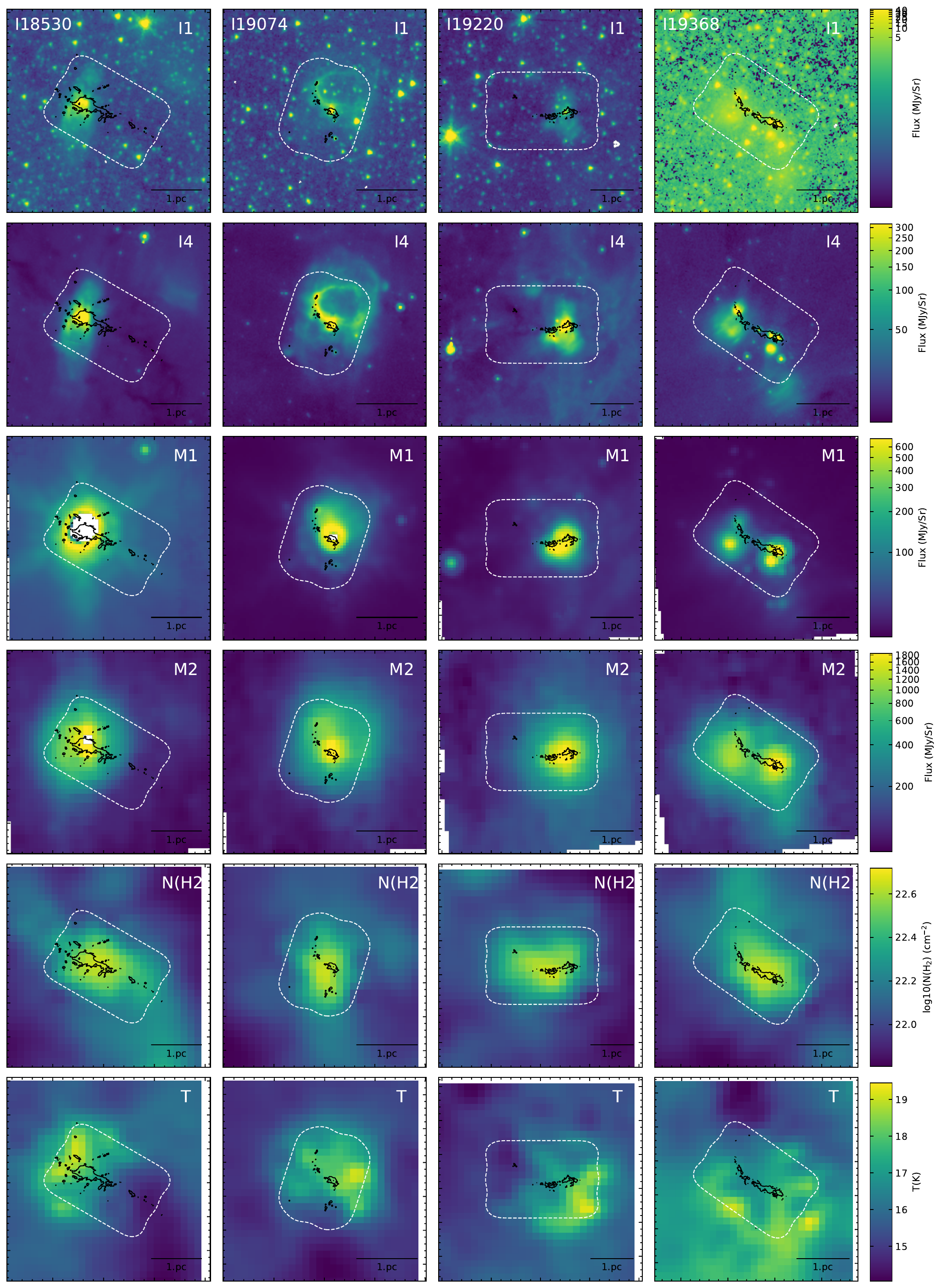}
\caption{Continued. From left to right we show I18530, I19074, I19220 and I19368.}
\end{figure*}




\section{Temperature derivation with the \ammonia{} lines} \label{sec:app_nh3}
\begin{figure*}[ht!]
\epsscale{1.1}\plotone{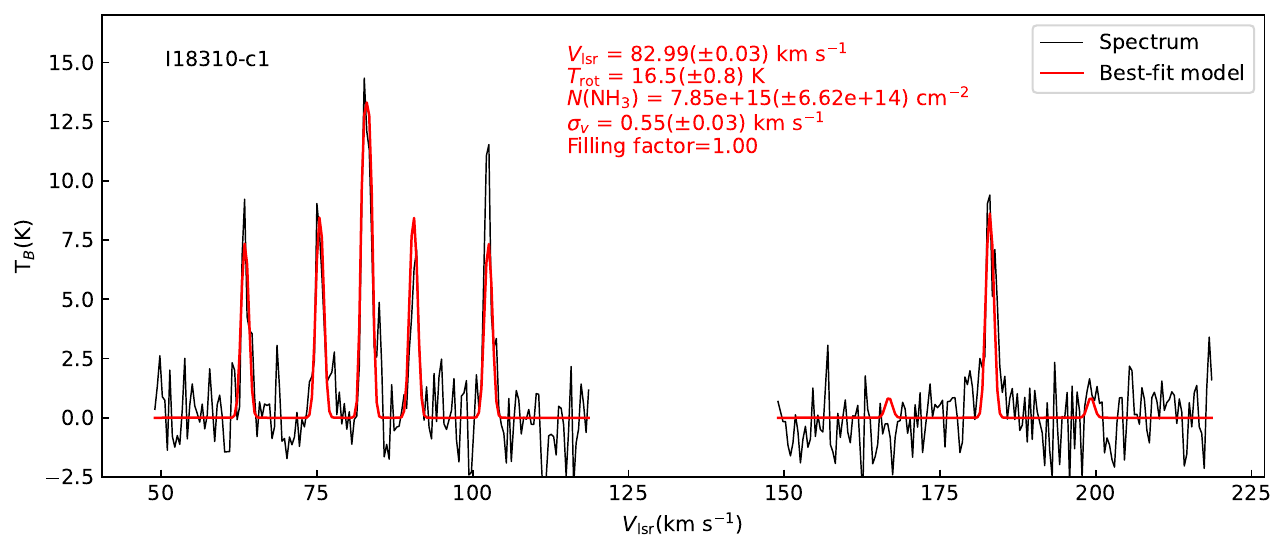}
\caption{ Fitting results of gas temperatures in I18310-c1 using the VLA \ammonia{} (1, 1) and (2, 2) lines. The \ammonia{}(2,2) line is manually shifted by 100~\kms. The red curve shows the best-fit model, with the best-fit parameters displayed in the panel.\label{fig:nh3_example}}
\end{figure*}

In \autoref{fig:nh3_example} we present an example of fitting \ammonia{} (1,1) and (2,2) lines with the routine in \citet{Lu18}. The errors displayed in the plot only account for the uncertainties in the fitting.

\section{Virial states of cores} \label{sec:app_virial}

\begin{figure*}[ht!]
\epsscale{1.0}\plotone{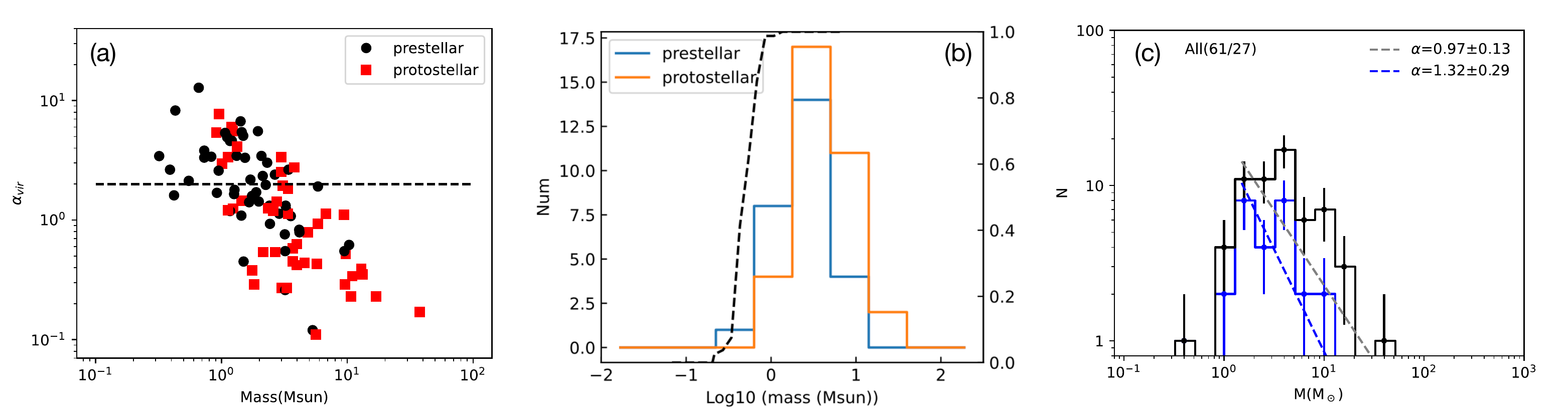}
\caption{ {\it (a)} Estimated virial parameters v.s. core masses. The prestellar and protostellar cores are shown in black and red, respectively. {\it (b)} Same as \autoref{fig:comp_mass}(d) but for the gravitionally bound population. { {\it (c)} Same as \autoref{fig:cmf}(a) but for the gravitionally bound population.}  \label{fig:virial}}
\end{figure*}

The virial parameters of dense cores can be estimated by 
\begin{equation}
\alpha = \frac{5 \sigma_{in}^2R}{GM_{\rm core}},
\end{equation}
where $\sigma_{\rm in}$ is the intrinsic 1D velocity dispersion of the molecule of mean mass and $R$ is the core radius \citep{Bertoldi92}. We use the velocity dispersion $\sigma_{\rm NH_3}$ of \ammonia{} lines to estimate $\sigma_{\rm in}$. This is done by first subtracting the channel width 0.4~\kms{} quadratically, i.e.,
\begin{equation}
\sigma_{\rm deconv,NH_3} = \sqrt{ \sigma_{\rm NH_3}^2 - (0.4~{\rm km\ s^{-1}}/2\sqrt{2{\rm ln2}} )^2   }   ,
\end{equation}
and then 
\begin{equation}
\sigma_{\rm in} =  \sqrt{\sigma_{\rm nth}^2+\sigma_{\rm th}^2} \\
             =  \sqrt{\sigma_{\rm deconv,NH_3}^2-\frac{k_BT_{\rm rot}}{\mu_{\rm NH_3} m_p}+\frac{k_BT_{\rm rot}}{\mu_p m_p}}.
\end{equation}
where $\mu_p$ = 2.33 is the mean molecular weight assuming $n_{\rm He}=0.1 n_{\rm H}$ and $\mu_{\rm NH_3}$ is the molecular weight of \ammonia. For the radius we adopt half of the deconvolved FWHM, i.e.,
\begin{equation}
{\rm FWHM_{ decov}} = \sqrt{{\rm FWHM_{ major}}\times {\rm FWHM_{ minor}} - {\rm Beam_{ major}} \times {\rm Beam_{ minor}}  } .
\end{equation}
We set a minimum deconvolved size of half the beam, to limit deconvolution effects that may give excessively small and thus unrealistic sizes. Thus the virial parameters are estimated for 104 cores with valid radius and velocity dispersion estimation. 
For a self-gravitating, unmagnetized core without rotation, a virial
parameter above a critical value $\alpha_{\rm cr}= 2$ (assuming a constant radial density profile) indicates that the core is unbound and may expand, while one below 2 
suggests that the core is bound and may collapse. In \autoref{fig:virial} we plot the virial parameters against the core mass. There is a trend for more massive cores having smaller virial parameters. 61 cores have virial parameters smaller than 2, suggesting they are gravitational bound. If we compare the mass distribution of prestellar and protostellar cores for the gravitational bound population, which is shown in \autoref{fig:virial}, we find the conclusion in \autoref{sec:core_growth} still holds, i.e., the protostellar cores are statistically more massive. The null hypothesis that protostellar cores have equally or smaller fluxes than prestellar cores can be rejected with a confidence greater than 99\% ($p$ = 0.0028).

\section{Comparison of CMFs derived by {\it astrodendro} and {\it getsf}}\label{sec:app_comp_method}

\begin{figure*}[ht!]
\epsscale{1.1}\plotone{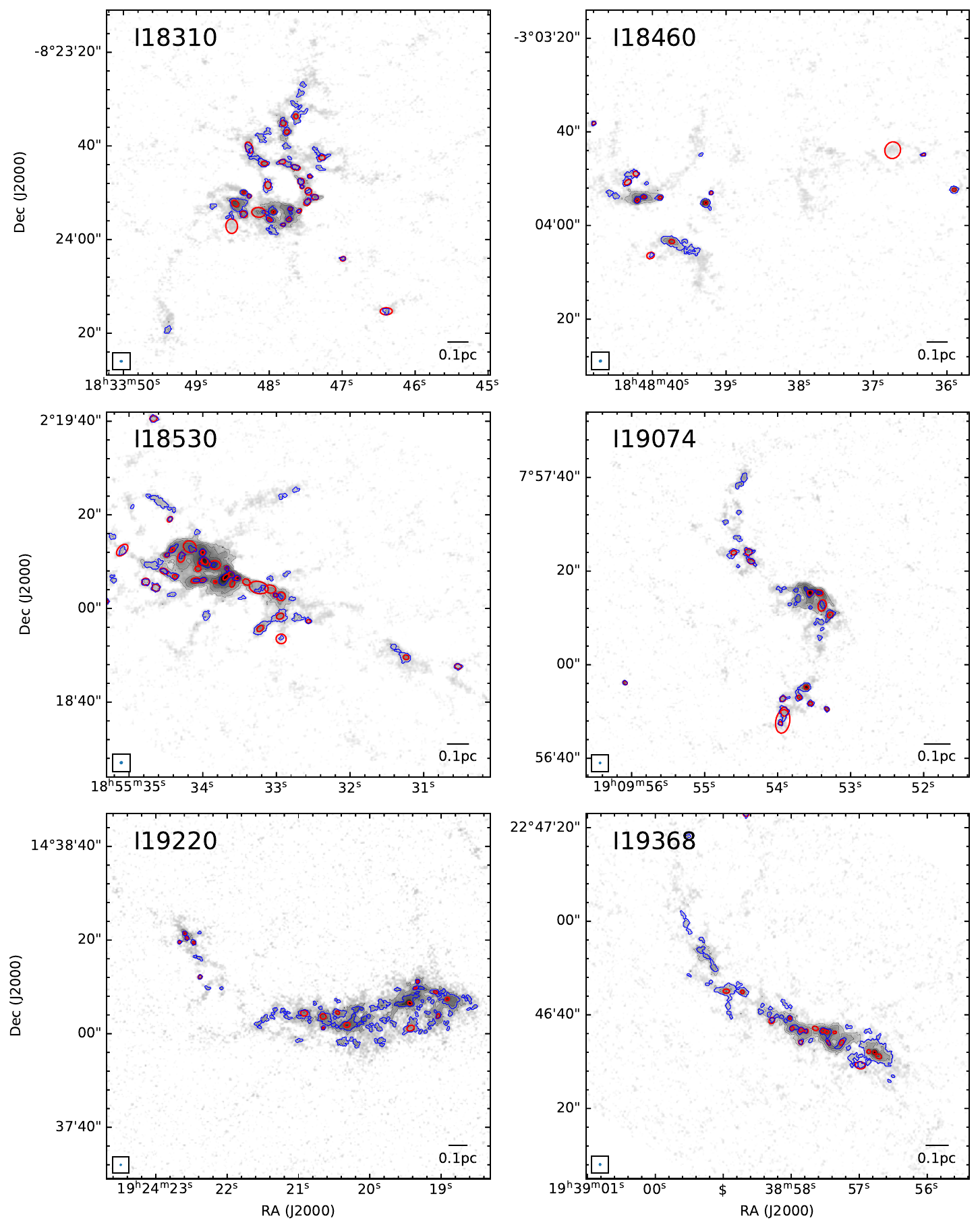}
\caption{Core identification results with {\it getsf} and {\it astrodendro}. Same as \autoref{fig:map_iden} but for other targets.\label{fig:map_iden_other}}
\end{figure*}

\begin{figure*}[ht!]
\epsscale{1.0}\plotone{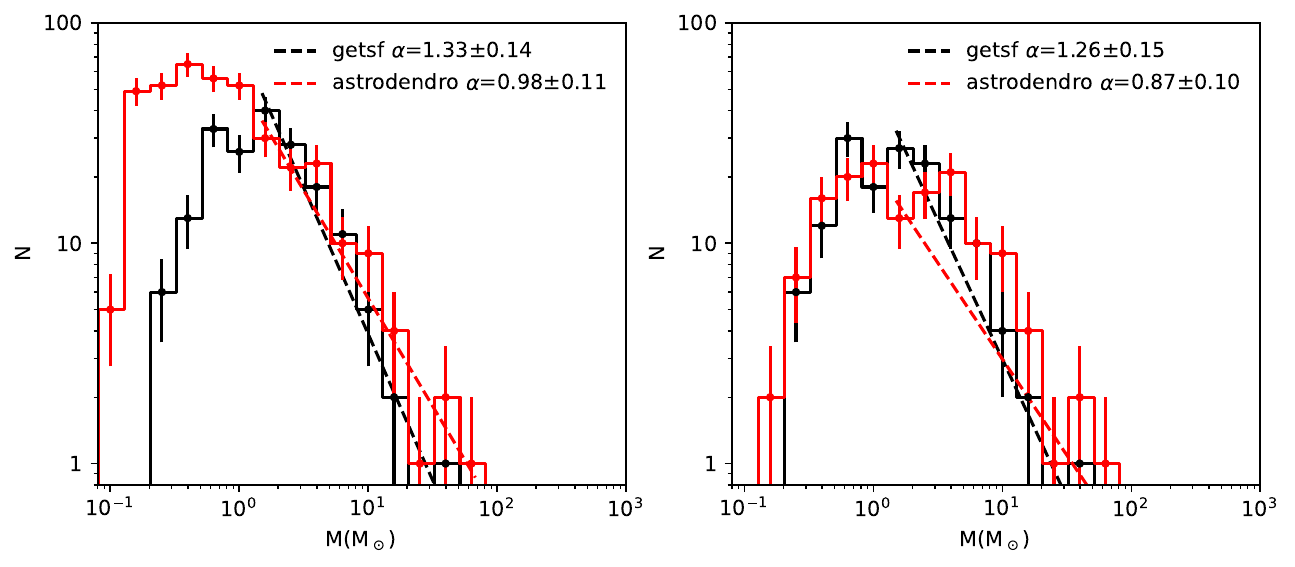}
\caption{{\it Left:} CMF results with {\it getsf} and {\it astrodendro} assuming a uniform temperature of 20~K. {\it Right:} Same as {\it Left} but only for cores commonly detected in both algorithms. The dashed lines indicate the power law fit by \autoref{equ:imf} for mass greater than 1.5~\msun.\label{fig:cmf_comp}}
\end{figure*}

\begin{figure*}[ht!]
\epsscale{1.0}\plotone{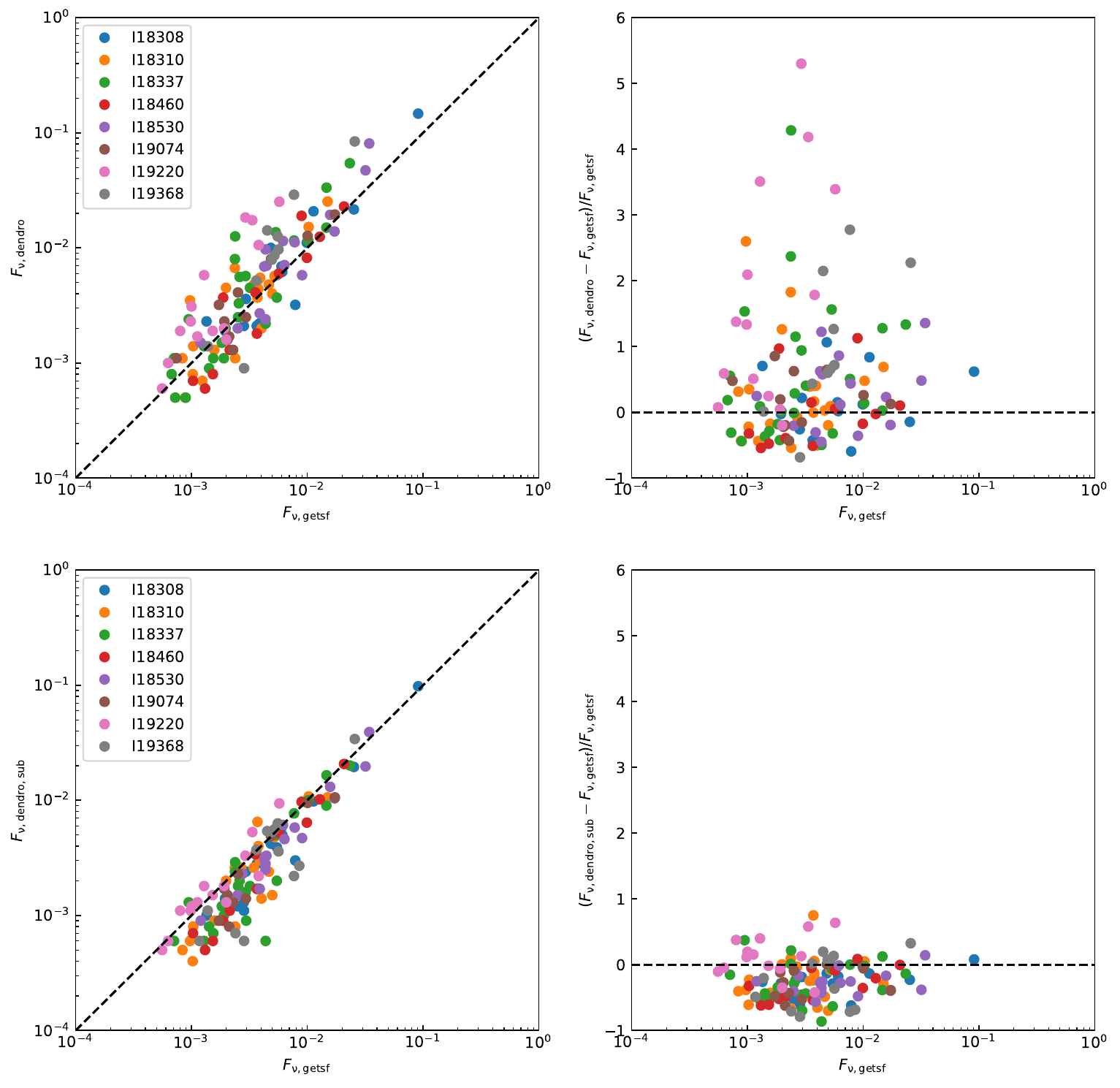}
\caption{Upper: Comparison of core flux densities returned by {\it getsf} and {\it astrodendro}. { Bottom: Same as upper panels but for {\it astrodendro} we removed the background component.} \label{fig:flux_comp}}
\end{figure*}

In \autoref{fig:map_iden_other} we show the core identification results with {\it astrodendro} and {\it getsf} for targets that are not shown in \autoref{fig:map_iden}. We present in \autoref{fig:cmf_comp} the CMFs derived with both algorithms, where the fluxes are converted to masses assuming a uniform temperature of 20~K. {\it Astrodendro} gives a larger number of cores (382) compared to {\it getsf} (183). A power law fit for the high mass end (M $>$ 1.5~\msun) gives a shallower slope $\alpha$=0.98$\pm$0.11 for {\it astrodendro}. We further check the CMF for 146 commonly detected cores in both algorithms, defined as a consistent position within 0\farcs{3}. The power law fit for the same range gives $\alpha$=0.87$\pm$0.10 for {\it astrodendro}, shallower than {\it getsf} ($\alpha$=1.26$\pm$0.15). But for {\it astrodendro} the CMF appears as a broken powerlaw with a separation point around $\sim$4~\msun. If the fit only accounts for M $>$ 4~\msun, then it gives $\alpha$=1.44$\pm$0.23. From inspection of this plot {\it astrodendro} appears to give more massive cores than {\it getsf} (e.g., M $>$ 4~\msun), and more low mass cores (e.g., M $<$ 0.8~\msun) but a deficit of cores in the intermediate mass range. We speculate the larger mass estimation of high mass cores for {\it astrodendro} is mainly due to the fact that these cores often lie on strong background emission and (at least in a few cases) {\it astrodendro} fails to identify the fragmentation in the extension wing of massive cores (see \autoref{fig:map_iden_other}). A comparison of the fluxes for commonly detected cores are shown in \autoref{fig:flux_comp}. It can be seen that the fluxes from two algorithms are correlated but the difference can be very large for individual cores, and {\it astrodendro} does return larger fluxes for the more massive cores. { For {\it astrodendro} we also attempted to remove the background using a median filtering method with a Gaussian Kernel of 5\arcsec. This mitigates the differences in fluxes and results in a better correlation between the fluxes obtained from the two methods.}

\section{Slopes of CMFs and cloud evolution}\label{sec:app_comp_method}

{ We test the robustness of the conclusion in \autoref{sec:cmf} by varying the classification of cloud evolution stages. In \autoref{table:slope_evo} we list the MLE power law fit results for different subgroups of the sample based on different classification schemes. Case 1 is the fiducial evolution classification adopted in \autoref{sec:cmf} by combining information from the $L/M$ ratio and other evolution indicators (see discussion \autoref{sec:evolution}). For case 2 we include all eight regions and classify the four regions (I18308, I18310, I18337, and I18460) as young group since a significant portion of these clouds are still infrared dark (for wavelengths up to 24~$\mu$m), and the rest four regions are classified as evolved group. For case 3 we split the eight regions into two groups purely based on their $L/M$ ratio listed in \autoref{table:info_sample}. For case 4 we selected from the fiducial case two sources with lowest $L/M$ ratios for the young group, and two sources with highest $L/M$ ratios for the evolved group, respectively. In these cases, the young group exhibits a high mass slope ranging from $\sim$0.9 to $\sim$1.1, while the evolved group has a slope from $\sim$1.4 to $\sim$1.6. In all cases the power-law index $\alpha$ for the evolved group is slightly steeper. For case 4 the $\alpha$ of two groups are consistent within 1$\sigma$ uncertainty mainly due to limited number statistics.  } 

\startlongtable
\begin{deluxetable*}{ccccc}
\tabletypesize{\scriptsize}
\renewcommand{\arraystretch}{1.0}
\tablecaption{Power-law index $\alpha$ (M$>$1.5~\msun) for subgroups of the INFANT sample\label{table:slope_evo}}
\tablehead{
\colhead{Classification} & \colhead{Young} & \colhead{$\alpha$}  & \colhead{Evolved} & \colhead{$\alpha$}  
}
\startdata
1 & I18308, I18337, I18460 & 0.89$\pm$0.15  &  I18530, I19220, I19074 & 1.44$\pm$0.25 \\ 
2 & I18308, I18310, I18337, I18460 & 1.01$\pm$0.14  &  I18530, I19220, I19074, I19368 & 1.44$\pm$0.21 \\
3 & I18308, I18337, I18460, I19368 &  1.00$\pm$0.14 & I18310,I18530, I19220, I19074   & 1.39$\pm$0.19\\
4 & I18337, I18460 & 1.06$\pm$0.23  &   I19220, I19074 & 1.55$\pm$0.49 \\
\enddata
\end{deluxetable*}

\bibliographystyle{aasjournal}
\bibliography{refer}

\end{CJK}
\end{document}